\documentclass{article}[affil-it]

\usepackage[left=2cm,right=4cm,top=1.5cm,bottom=1.5cm]{geometry}
\usepackage[utf8]{inputenc}
\usepackage{outlines}
\usepackage{svg}
\usepackage{soul}
\usepackage{amssymb, bm}
\usepackage{pgfplots}
\usepackage{eurosym}
\usepackage{comment}
\usepackage{mathtools}
\DeclarePairedDelimiter{\ceil}{\lceil}{\rceil}
\usepackage{carshapes}
\usepackage{multirow}
\usepackage{amsfonts}
\usepackage{comment}
\usepackage{booktabs}
\usepackage{siunitx}
\usepackage{xcolor}
\usepackage{float}
\usepackage[ruled,vlined, noend]{algorithm2e}
\usepackage{caption}
\usepackage{subcaption}
\usepackage{amsmath}
\usepackage{enumitem}
\usepackage{algorithmicx}
\usepackage{algcompatible}

\usepackage{tikz}
\usetikzlibrary{shapes.multipart}
\usetikzlibrary{arrows, chains, positioning, shapes.geometric, shapes.symbols}
\usepackage[draft]{todonotes}
\usepackage{amsopn}
\usepackage{breakcites}
\usepackage[affil-it]{authblk}
\algrenewcommand{\algorithmiccomment}[1]{\footnotesize {\fontfamily{ppl}\selectfont \% #1}}
\usetikzlibrary{arrows.meta}
\pgfplotsset{compat=1.16}

\title{Anticipatory routing methods for an on-demand ridepooling mobility system }
\author[1]{Andres Fielbaum
  \thanks{a.s.fielbaumschnitzler@tudelft.nl; Corresponding author}}
  \author[1]{Maximilian Kronm\"uller}

  \author[1]{Javier Alonso-Mora}
  \date{}
\affil[1]{Department of Cognitive Robotics, TU Delft, The Netherlands}

\begin{document}

\maketitle
\abstract{On-demand mobility systems in which passengers use the same vehicle simultaneously are a promising transport mode, yet difficult to control. One of the most relevant challenges relates to the spatial imbalances of the demand, which induce a mismatch between the position of the vehicles and the origins of the emerging requests. Most ridepooling models face this problem through rebalancing methods only, i.e., moving idle vehicles towards areas with high rejections rate, which is done independently from routing and vehicle-to-orders assignments, so that vehicles serving passengers (a large portion of the total fleet) remain unaffected. This paper introduces two types of techniques for anticipatory routing that affect how vehicles are assigned to users and how to route vehicles to serve such users, so that the whole operation of the system is modified to reach more efficient states for future requests. Both techniques do not require any assumption or exogenous knowledge about the future demand, as they depend only on current and recent requests. Firstly, we introduce rewards that reduce the cost of an assignment between a vehicle and a group of passengers if the vehicle gets routed towards a high-demand zone. Secondly, we include a small set of artificial requests, whose request times are in the near future and whose origins are sampled from a probability distribution that mimics observed generation rates. These artificial requests are to be assigned together with the real requests. We propose, formally discuss and experimentally evaluate several formulations for both approaches.

We test these techniques in combination with a state-of-the-art trip-vehicle assignment method, using a set of real rides from Manhattan. Introducing rewards can diminish the rejection rate to about nine-tenths of its original value. On the other hand, including future requests can reduce users' traveling times by about one-fifth, but increasing rejections. Both methods increase the vehicles-hour-traveled by about 10\%. Spatial analysis reveals that vehicles are indeed moved towards the most demanded areas, such that the reduction in rejections rate is achieved mostly there.}

\section{Introduction}

Centrally controlled on-demand ridepooling systems, in which different users can ride the same vehicle at the same time if their paths are compatible, are a promising mobility system for the future of cities, because they can exhibit many of the advantages of popular (non-shared) on-demand systems more sustainably without increasing congestion. 

Massive on-demand systems (apps) have become popular due to a number of virtues: short waiting times, door-to-door service, ease of payment, an increase of comfort, and no need for parking nor driving \cite{rayle2016just,tirachini2019ride,tang2019app}. All these positive features can be kept when rides are shared (pooled) as well.

Moreover, sharing can effectively fight congestion and emissions if an adequate fleet is selected \cite{tirachini2020sustainability,li2021does}. Empirical studies have shown that carsharing systems in which rides are not shared have increased congestion, as they attract many users from public transport \cite{henao2019UberVMT,tirachini2019UberVKT,agarwal2019impact,roy2020traffic,diao2021impacts,wu2021assessing}. When rides are shared, vehicles make more efficient use of the scarce vial space.

However, such mobility systems are quite hard to operate, as they combine the principles of two classic NP-Hard problems: the Dynamic-Vehicle-Routing-Problem (as they are on-demand) and the Dial-A-Ride-Problem (due to the sharing aspect). Despite this complexity, some algorithms have been able to effectively decide how to match groups of users and how to assign them to vehicles \cite{ota2016stars,alonso-mora_-demand_2017,tsao2019model,simonetto2019real,kucharski2020exact}, and their simulations have confirmed the potential of ridepooling.

\begin{figure}[H]
    \centering
\includegraphics[width=6.5cm]{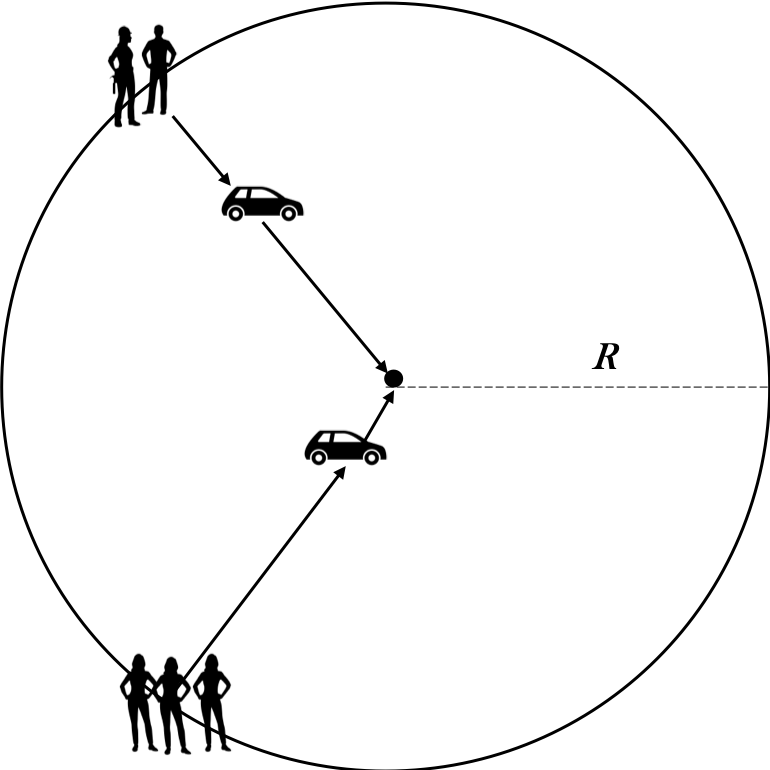} 
\caption{A ridepooling system in an extreme version of a morning peak situation. If the time required to drive $R$ exceeds the maximum waiting time, all vehicles will get stuck in the center.}
\label{Img:CircularCity}
\end{figure}

One of the main difficulties of massive on-demand systems is related to their dynamics. The system needs to decide the assignments as the requests appear. Even if these assignments are decided optimally according to the current conditions, they might leave the system in a state that is inefficient to serve the demand that will emerge afterward. Let us consider an extreme and simplified example that helps to visualize the situation: a circular city, in which the users are located at the border and  are all traveling to the center. Figure \ref{Img:CircularCity} shows such a scenario, simulating an exaggerated version of a morning peak situation. This demand unbalanced demand pattern will make all the cars converge rapidly at the center; if there are bounds on the maximum waiting time (a usual assumption in these models) and the time required to go back to the border exceeds this bound, then vehicles will not be assigned to any new request, and the system would collapse.

To prevent such situations, most of the algorithms include a \emph{rebalancing} step, in which cars \emph{that are not being used} are sent to some zone in which they are expected to be required. Note that underlying this idea, there has to be some way to determine which zones of a city will require vehicles in the future, i.e., there are some \emph{anticipatory} decisions. The corresponding scheme of decisions can be synthesized in the following steps:

\begin{enumerate}
    \item One or more trips are requested to the system.
    \item The system decides how to assign the request(s) to the set of vehicles, i.e., which new users will travel together, in which vehicle(s) and following which route. This process is performed according to some optimization rules and constraints that deal with the quality of service. Sometimes it is not possible to serve all the requests, and some get rejected.
    \item After assigning or after some predefined lapse of time, the system determines which vehicles are not being used and decides how to rebalance them.
\end{enumerate}

Two relevant aspects underlie the recent description: rebalancing does not interact directly with assignments, and it only deals with idle vehicles. These aspects can have substantial drawbacks. A relevant limit to rebalancing strategies is given by the number of vehicles that can be controlled, which might be much lower than the total number of vehicles in the system when only idle vehicles are considered. Such a situation is particularly relevant when the system is serving close to its maximum capacity, in which case most vehicles are being used constantly and thus are not available for rebalancing. This is quite problematic as these heavily loaded scenarios are the ones in which the system needs to work as efficiently as possible. 

Furthermore, sometimes the assignment process may produce some inefficient matches, which cannot be corrected through rebalancing. In particular, some vehicles might be directed towards low-demand areas to give a better quality of service at a specific time, but being barely shared. Such situations worsen future service, as vehicles move to zones in which they are not likely to be required or used to their full potential (if the remain being low-demand).

\begin{figure}

    \centering
    \qquad
\subfloat{{\includegraphics[width=4cm]{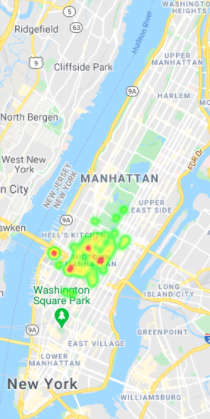} }}
   \qquad
\subfloat{{\includegraphics[width=3.95cm]{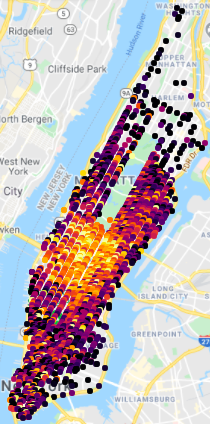} }}
\qquad
\subfloat{{\includegraphics[width=4cm]{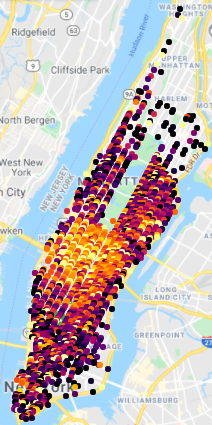} }}
\caption{Left: Rejection rate (the more red the higher, and no color means almost no rejections). Center: average waiting time (the clearer the higher). Right: average delay (the clearer the higher). All three are obtained for trips departing at each node in Manhattan's network, after simulating one hour of a ridepooling system, just after midday on 15/01/2013. They all show that the most demanded zone, at the center of the network, faces a worse quality of service. Source: \cite{fielbaum2021demand}}
\label{Img:PaperWalking}
\end{figure}

These phenomena can be observed in Figure \ref{Img:PaperWalking} (from \cite{fielbaum2021demand}), which shows the quality of service when modeling a ridepooling system (based on the model by \cite{alonso-mora_-demand_2017}) during one hour in Manhattan. It reveals that the level of rejected requests, the average waiting time, and the average delay are all higher at the center of the network, precisely the most demanded zone. Similarly,  \cite{alonso-mora_-demand_2017} provides a video\footnote{A full version of the video is available at https://youtu.be/xHWrRci0H54. Accessed: 29/06/2020.} that shows the daily performance of the system. In this video, two facts are noteworthy: there are almost no rebalancing vehicles, and the number of passengers per vehicle is much higher in the most demanded zone than in the rest of the network.

These problems are inherent to the system's combination between sharing and having flexible routes. When cars are shared but rides are not (i.e., on-demand taxis), finding a single passenger is as good as possible, so there is no need to go to the most demanded zones if there is some demand in the rest of the network. In public transport, lines are designed a priori to serve better the most demanded zones through higher frequencies, shorter distances to bus stops, and more direct services \cite{mohring1972optimization,chang1991multiple,daganzo2010structure,fielbaum2020beyond}, i.e., the mismatch between supply and demand is prevented thanks to having fixed routes. 

To prevent such situations, techniques that are beyond rebalancing are needed. In this paper, we study methods that introduce anticipatory decisions at an earlier stage, namely when deciding the assignments (i.e., which vehicle is carrying which passengers) and the routes (i.e., in which order are they served), in order to have the whole system better prepared for future service. The techniques we propose do not require any exogenous assumptions about future requests. They take as input only the endogenous information that is generated while operating the system, namely the origins of current and recent requests (we also study the use of historical data as a benchmark, and we show it performs worse). As revealed by our simulations based on real-life data, current and recent requests can be used as an efficient proxy for what will happen in the near future when applying our techniques. Moreover, by these means, our techniques adapt endogenously to the different situations in which they might be applied.

We study two types of techniques: First, by modifying the cost of each possible assignment between vehicles and set of requests, favoring those assignments that conduct the vehicle towards the most demanded zones; second, we introduce artificial future requests to the pool of requests that have to be assigned at each time, with origins in those same high-demanded zones so some vehicles might be sent towards them. Several specific implementations for each technique are analyzed and compared.

The paper is structured as follows: Section \ref{relatedResearch} synthesizes previous research on the topic and highlights the contributions of this paper over the state of the art. Section \ref{scn:methods3} explains the two main techniques that we propose in this paper, which depend on some ``generation'' and ``rejection'' rates, that are explained in section \ref{scn:rates4}. In section  \ref{Scn:ResultsPaperAnt6} we run simulations to analyze the impact of the methods over a real-life case (in Manhattan), together with a deep analysis of how the spatial mismatch between vehicles, requests, and rejections is modified. Finally, in section \ref{scn:conclusionsAnt7} we conclude and propose several directions for future research.

\section{Related Works} \label{relatedResearch}
\subsection{On-demand ridepooling systems}
The research over on-demand ridepooling systems has rapidly expanded during the last years, boosted by the success of on-demand mobility apps (both shared and non-shared) and the need for more sustainable systems in which vehicles are used for more than one passenger. Many studies deal with the problem of how to assign passengers and vehicles, either through agent-based models (like \cite{merlin2017comparing,fagnant2018dynamic,lokhandwala2018dynamic,vosooghi2019shared} among others) or through centralized algorithms (like \cite{d2012empirical,pelzer2015partition,alonso-mora_-demand_2017,gao2017efficient,qian2017optimal,wang2018stable,tsao2019model,simonetto2019real,fielbaum2021demand} among others). Some of these assignment methods have rebalancing or anticipatory techniques, which are described in the following subsection. 

These systems have been further analyzed in a number of other directions. A relevant issue related to this paper is that on-demand ridepooling can be more unreliable (i.e., it is harder for users to predict which quality of service they will experience) than other mobility systems. This happens due to the same dynamics that call for anticipatory techniques, i.e., flexible routes that respond to the emerging demand. These unreliability sources are described in detail by \cite{fielbaum2020unreliability}, which focus on those sources related to vehicles' assignments. On a related note, \cite{Kucharski2020If,hyland2020operational} study the impact of late passengers, while \cite{liu2019dynamic} analyzes the impact of traffic congestion, all of which affect the future performance of the system.

\subsection{Rebalancing and predictive methods} \label{scn:related}
Many of the ridepooling\footnote{Many rebalancing methods have been proposed for non-shared systems: \cite{yu2019markov,braverman2019empty,zhang2016control,pavone_robotic_2012,gao2018optimize}, among others. We are not studying them in detail here because there are fundamental differences when sharing is admitted. In the non-shared case, finding a single request as fast as possible is the best that can be done, whereas here, we aim to maximize the number of requests that a vehicle can find soon. In this last case, it is much more relevant to send the rebalancing vehicles to high-demand areas instead of zones in which is likely to find at least one request.}
systems proposed in the literature include some rebalancing method that decides how to move idle vehicles. The general idea underlying all of them is the following: if a vehicle is idle right now, the place in which it is currently located has an oversupply of vehicles, so it should move to a different location where it is more likely to be required. The decisions of which vehicles to move and towards which zones may also try to prevent the rebalancing vehicles to drive for too long distances.

The methods present in the literature differ mostly on how to measure the need for vehicles in each region. Some papers consider the current demand \cite{vosooghi2019shared,lioris2016dynamic}; \cite{spieser_toward_2014} seeks for an even distribution of the vehicles; \cite{alonso-mora_-demand_2017} (the model we use here for the simulations, and whose rebalancer we also include) takes the origin of the rejected requests as the mark that defines that a vehicle is needed there, an approach that is adapted by \cite{liu2019proactive}, who excludes from the rebalancing process vehicles that were already being rebalanced, and by \cite{tsao2019model}, who  keeps some predefined number of vehicles idle to be able to serve unexpected demand;  \cite{van2018enhancing} assumes that the probabilities of having new trips in a zone are known, based on historical data, and compares it with the current number of vehicles that are currently in such zone; and \cite{wallar2018vehicle} estimates the future demand using a particle filter method that considers the temporal evolution of the demand within each zone. Finally, \cite{wen2017rebalancing} uses an approach that is different from the ones explained so far, as they use reinforcement learning techniques for rebalancing, which they show to outperform a rebalancer similar to the ones previously described, and a simpler rebalancer that randomly moves a vehicle to its neighboring areas, with probabilities proportional to their demand rates. As discussed in the introduction, all these rebalancing methods affect idle vehicles only.

Few papers deal with anticipatory routing or assignments, as we do here. In a simplified context, in which users travel between specific stations among the area covered by the system, \cite{barth2004user} propose a method that could be used in more general schemes: splitting groups of passengers into many vehicles if their destinations are located in a zone that is expected to require more vehicles in the near future, thus routing more vehicles into those areas. \cite{van2018enhancing} works with an event-based model, in which each time a new request emerges, it is assigned to a vehicle that is chosen using demand forecast, trying to match the number of empty seats of that vehicle with the number of users that it might pick up when transporting the original request; similar ideas are proposed by \cite{wang2020demand}, but assigning several requests at a time.

Two papers propose predictive routing and assignment ideas building upon the same model that we use for simulations (\cite{alonso-mora_-demand_2017}), and using related techniques: \cite{alonso-mora_predictive_2017}) estimates historical demand for each zone, and includes some artificial requests according to this estimated demand in order to push the whole system towards a better preparation for the future, an approach that we extend here considering alternative ways of defining the artificial requests, other than based on historical data; \cite{huang2018efficient} modifies the cost function with an additive term that depends on the spatial distribution of the vehicles and its distance to the optimal one. These papers have some drawbacks: the former requires historical data that is not always available, it increases the computational time heavily, and it does not have a meaningful impact on reducing the number of rejected requests of the system (it does reduce average waiting times and delay); while the latter requires a perfect knowledge of the demand distribution, and it does not affect the routes of the vehicles for a given set of pick-ups and drop-offs.

In all, there is a vast literature showing that rebalancing techniques for idle vehicles can have significant impacts on the efficiency of ridepooling systems, but just a few papers that deal with the assignments and routes for vehicles that are actually being utilized, which might be the majority within the fleet. Such papers work over specific assignment methods, and usually require some knowledge either about historical data or future demand, instead of leveraging the information generated endogenously by the system when operated.

\subsection{Contribution}
The contributions of this paper are threefold:
\begin{enumerate}
    \item We propose two new methods to modify the decision scheme of a ridepooling system, aiming to face future requests with a better quality of service. Unlike previous works, these methods affect the system's decision process at every stage and are compatible with different assignment procedures. The theoretical virtues and drawbacks of both methods are discussed in detail.
    \item We propose several ways to define generation and rejection rates per zone in the city, which are required by the anticipatory methods but can be used for other purposes as well. Such rates are computed using data that is endogenously generated when operating the system, so that no exogenous information or assumptions are required.
    \item We run detailed experiments to analyze under which conditions the methods improve the system's performance, and propose various ways and metrics to analyze how vehicles' routes are modified, allowing us to measure whether the methods push the system in the expected direction. Results show that the methods are able to improve the system, and that they do so by reducing the mismatch between the positions of vehicles and requests, i.e., the methods effectively allocate more vehicles in zones where they are needed to serve their large demand.
\end{enumerate}

\section{Two Anticipatory Methods} \label{scn:methods3}
This paper proposes two anticipatory methods to modify already existing algorithms that decide how to assign requests and vehicles in an on-demand ridepooling system. The first one (``introducing rewards'') modifies how to value each possible route for a vehicle, impacting which route to select and how to assign vehicles to requests. The second method (``insertion of artificial requests'') includes future trips to the list of requests to be served, which imposes a heavier computational burden, but with the virtue of acting through a global evaluation of the system, moving some vehicles towards the origins of such future requests. 

To explain both methods in detail, we first introduce some terminology and the formal statement of the assignment problem.
\subsection{Problem statement} \label{scn:statementanticipatory}

We aim to match requests together and assign them to vehicles efficiently, during some predefined period of operation that lasts $PO$. However, the requests that are going to be served are not known beforehand but emerge throughout the operation. Therefore, this problem belongs to the well-studied family of \textit{stochastic optimization}, as surveyed by \cite{powell2019unified}, although we shall not assume any probability distribution for the unknown future events.

\subsubsection{Terminology}
Let us first introduce relevant terminology and notation\footnote{A glossary with all the math symbols used throughout the paper is provided in the appendix.}:
\begin{itemize}
    \item The problem takes place over a directed graph $G=(V,E)$ representing the road network used by the vehicles.
    \item A \emph{request} $r$ is a single call from a user (or a number of them traveling together) that emerges at time $tr_r$, and needs to be transported from an origin $o_r$ to a destination $d_r$, both located on the nodes of the graph. The set of all requests that emerge during the period of operation is denoted $\mathcal{R}_{all}$. Requests that have emerged up to time $t$ (i.e., such that $tr_r \leq t$) can be divided into four sets: $\mathcal{R}_s(t), \mathcal{R}_e(t),\mathcal{R}_c(t)$ and $\mathcal{R}_r(t)$ representing, respectively, those that are being served by the system (either waiting for an assigned vehicle or en-route), those that have recently emerged and are waiting to be assigned, those that are completed and those that were rejected by the system. Note that, at the end of the operation ($t=PO$), it must hold that $\mathcal{R}_{all}= \mathcal{R}_s(t) \cup \mathcal{R}_e(t) \cup \mathcal{R}_c(t)  \cup \mathcal{R}_r(t)$. 
    \item A \emph{vehicle} $v$ is characterized at each time $t$ by its capacity $\mu_v$, its current position $Pos_v(t)$, a set of requests that it is currently serving $Req_v(t)$, and its planned route $\pi_v(t)$ (i.e., a path over the graph). The set of vehicles at time $t$ is denoted $\mathcal{V}(t)$.
     \item A \emph{trip} $T=(r_1,...,r_k)$ is a set of requests in $\mathcal{R}_e(t)$. Such a trip is feasible to be transported by a vehicle $v$ at time $t$, if there exists a route $\pi$ that serves the requests in $T$ and in $Req_v(t)$, fulfilling a set of constraints denoted by $\mathcal{C}$. Usual constraints include maximum waiting times and delay for each request, the vehicles' capacities, and can consider some additional rules like FIFO (first-in-first-out).
    \item Consider a feasible matching between a vehicle $v$ and a trip $T$ at time $t$, that instructs $v$ to follow a new route $\pi$. We define the cost $c(v,T,\pi,t)$ induced to the system, that might include the costs for requests in $T$, extra costs for requests in $Req_v(t)$ (because the updated route might induce longer traveling times for such requests), and operator's costs $c_O$. Users' costs $c_U$ usually depend on the time they spend waiting to be picked up by a vehicle and the time spent en-route, and might also take into account other features such as the number of co-travelers or the discomfort induced by new changes on the vehicle's route \cite{fielbaum2020unreliability}.
\end{itemize}

\subsubsection{A problem that is decided over time with partial information}
As this is a dynamic problem, there will be a set of time instants $\Psi=\{\tau_1,...,\tau_N\}$, such that at each instant $\tau_i$ the system decides an \textit{assignment} between $\mathcal{V}(\tau_i)$ and $\mathcal{R}_e(\tau_i)$, i.e., the system allocates the requests in $\mathcal{R}_e(\tau_i)$ (the ones waiting to be assigned) to the fleet of vehicles. How to define $\Psi$ is a decision of the operator of the system. If the decisions are taken too often, the system will not collect much information each time, which might yield inefficient decisions; on the other hand, if there is a long lapse of time between two consecutive decisions, some requests will have to wait too long before getting assigned to a vehicle, yielding a bad quality of service.

Two usual approaches for defining $\Psi$ are i) event-based ones, in which each time a request emerges it is immediately assigned to a vehicle (like \cite{van2018enhancing}), i.e., $\Psi=\{tr_r: r \in \mathcal{R}_{all} \}$ and ii) to perform an assignment process each fixed $\delta$ (like \cite{alonso-mora_-demand_2017}), i.e., $\Psi=\{k \cdot \delta: k = 1,...,\ceil{PO/\delta} \}$. The first approach could be easily extended to assign each time a fixed number of $b$ requests have emerged. According to the terminology discussed by \cite{powell2019unified}, the first approach corresponds to an online algorithm, while the second one is a receding horizon (or model predictive control, or rolling horizon). Both ideas share that \textbf{no knowledge about the future is assumed, a characteristic that we assume for the rest of the paper}. Instead, we will exploit the fact that there is some correlation between recent and upcoming demand.

Consider a time $\tau_i \in \Psi$. Formally, an assignment $A$ decided at that time is a set of vehicle-trip-route tuples $(v,T,\pi)$, so that $v$ is told to serve $T$ following the route $\pi$, with $T \subseteq \mathcal{R}_e(\tau_i)$.  An assignment is feasible if every matching between a vehicle and a trip fulfills the constraints $\mathcal{C}$, every vehicle is assigned to no more than one trip, and every request is assigned to no more than one vehicle. Note that some requests might not be assigned to any vehicle, forming the set of rejected requests $Q(\tau_i,A)$. An assignment might include \textit{reassignments}, i.e., updating some decisions taken before; for instance, it could modify the vehicle assigned to a request that has not been picked up yet, which is formally done by keeping those requests in $\mathcal{R}_e(\tau_i)$. 

The set of all possible feasible assignments at time $\tau_i$ is denoted by $\mathcal{A}(\mathcal{R}_e(\tau_i),\mathcal{V}(\tau_i),\mathcal{C})$. Each assignment $A$ has a total cost that depends on the specific costs $c(v,T,\pi,\tau_i)$ for each tuple $(v,T,\pi) \in A$, and a penalty for the rejections $c_R(Q(\tau_i,A))$ (for instance, proportional to the size of $Q(\tau_i,A)$). Once an assignment has been decided, each vehicle's information is updated, as well as the status of each request (i.e., to which $\mathcal{R}_{\bullet}(\tau_{i+1})$ they belong).
\subsubsection{The assignment problem over the period of operation}

Putting everything together, we can now formally define the assignment problem. With no future information\footnote{Other papers assume some probability distribution regarding future requests, so their approach is optimizing the expected value of the corresponding objective function.}, the assignment problem consists in solving Eq. (\ref{Eq:assignment}) iteratively at each $\tau_i$:

\begin{equation} \label{Eq:assignment}
    \min_{A \in \mathcal{A}(\mathcal{R}_e(\tau_i),\mathcal{V}(\tau_i),\mathcal{C})} \sum_{(v,T,\pi) \in A} c(v,T,\pi,\tau_i) + c_R(Q(\tau_i,A))
\end{equation}

Intuitively, Eq. (\ref{Eq:assignment}) selects an assignment between the requests to be assigned and the vehicles, minimizing the total costs of the system, which is determined by the costs faced by each trip-vehicle assignment and the penalty due to the rejected requests.

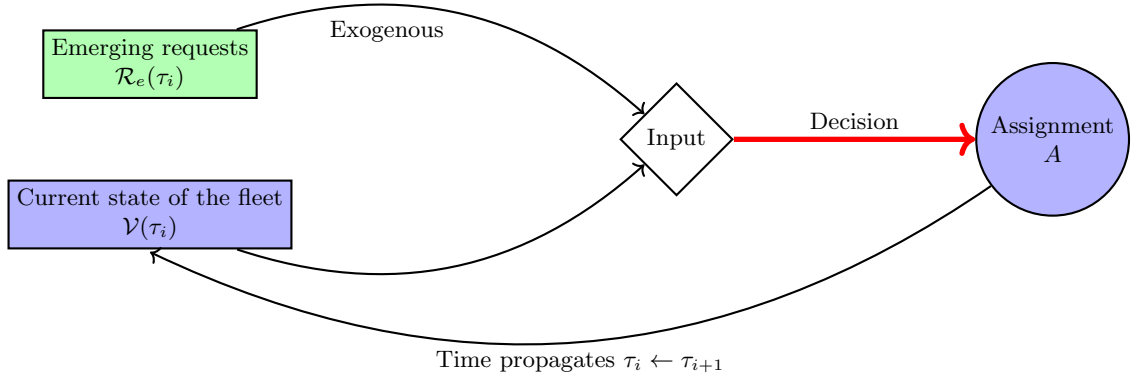
\begin{figure}[H]
    \centering
    \small
    \begin{tikzpicture}[->>=1pt,auto,node distance=1.5cm,
          thick,b node/.style={circle,draw},a node/.style={draw},c node/.style={diamond,draw},every text node part/.style={align=center}]

          \node[a node,fill=blue!30] at(-4,0) (1) {Current state of the fleet \\ $\mathcal{V}(\tau_i)$};
           \node[a node, fill=green!30] at(-4,2) (2) {Emerging requests\\$\mathcal{R}_e(\tau_i)$};
                     
           \node[c node, ] at(3,1) (3) {Input};
           
           \node[b node, fill=blue!30] at(8,1) (4) {Assignment \\ $A$};
           

          \path [->, bend right] 
           (1) edge node {} (3);
           \path [->, bend left,below left] 
            (2) edge node {Exogenous} (3);
            \path [->, line width=2pt,color=red] 
            (3) edge node {\color{black}Decision} (4);
            \path [->, bend left,below] 
           (4) edge node {Time propagates $\tau_i \leftarrow \tau_{i+1}$} (-4,-0.5);

        \end{tikzpicture}
    \caption{A diagram that synthesizes the assignment process during the whole period of operation. Exogenous requests emerge and need to be assigned to the available fleet of vehicles. The anticipatory techniques we introduce in this paper take place during the assignment decision (marked with a red arrow). Then time propagates forward (marked by the dotted arrow), and the fleet's status is updated until new requests emerge and the following assignment occurs. The green background represents exogenous inputs, whereas the blue background represents what is decided by the system.}
    \label{fig:DiagramAnticipatory}
 \end{figure}

The whole process is synthesized in Figure \ref{fig:DiagramAnticipatory}. At each stage (i.e., at each $\tau_i \in \Psi$), the requests in $\mathcal{R}_e$ have to be assigned to the current status of the vehicles. The anticipatory techniques we propose in this paper alter how to choose the optimal assignment, represented by the thick red arrow in Figure \ref{fig:DiagramAnticipatory}. This new assignment (with or without anticipatory techniques) updates the vehicles' routes, which defines the new status of the system until a new set of requests is inputted into the system, and a new stage begins.

Determining the feasible assignments $\mathcal{A}(\mathcal{R}_e(\tau_i),\mathcal{V}(\tau_i),\mathcal{C})$, as well as solving Eq. (\ref{Eq:assignment}), can be challenging from a combinatorial point of view, so some heuristics might be applied throughout the process, yielding to approximate solutions.

For synthesis, to decide how to form the groups and how to assign the vehicles, a method needs to define: $\Psi$, the cost functions, the constraints, and the techniques to compute $\mathcal{A}(\mathcal{R}_{e}(\tau_i),\mathcal{V}(\tau_i),\mathcal{C})$ and to solve Eq. (\ref{Eq:assignment}). We encompass all of this with the concept of \textbf{assignment procedure} $\mathcal{P}$. This concept is useful, because the anticipatory techniques will work on top of a predefined assignment procedure (for instance, when we run numerical simulations we will use the assignment procedure by \cite{alonso-mora_-demand_2017}).

In sections \ref{scn:rewards} and \ref{scn:artif} we propose techniques that affect how to solve one stage of the system, i.e., how to compute the assignments for a fixed $\tau_i$. The idea is to obtain a feasible assignment in $\mathcal{A}(\mathcal{R}_e(\tau_i),\mathcal{V}(\tau_i),\mathcal{C})$, whose cost might be higher than the optimal, but that yields better results in subsequent stages. The purpose is to reduce total costs at the end of the operation, i.e., to minimize the \textbf{a-posteriori-cost-function} given by Eq. (\ref{Eq:APosteriori}). There, $\mathcal{R}_{ok}$ is the set of all the requests that were served at the end of the operation, $\mathcal{R}_{ko}$ are those that were rejected, and $\Pi_{v}$ is the plan followed by $v$ through the operation; $c_U,c_R$ and $c_O$ represent the costs of the served users, the rejected users and the operator, respectively. It is worth noting that Eq. (\ref{Eq:APosteriori}) cannot be computed by adding up the value reached by Eq. (\ref{Eq:assignment}) at every $\tau_i$, because some of the requests that are assigned at $\tau_i$ might be reassigned to a different vehicle (or even rejected) at future assignments, which is why we need to compute everything a posteriori.

\begin{equation} \label{Eq:APosteriori}
    \sum_{r \in \mathcal{R}_{ok}} c_U(r) + c_R(\mathcal{R}_{ko}) + \sum_{v} c_O(\Pi_v)
\end{equation}

As the demand is not known beforehand, this function cannot be optimized a priori, but it is the underlying objective of the whole process, so it provides the benchmark to determine if the anticipatory techniques do improve the system's performance.

In sections \ref{scn:rewards} and \ref{scn:artif}, we assume that we have an assignment procedure. That is to say, we propose anticipatory techniques for generic definitions of the set $\Psi$, constraints $\mathcal{C}$ and cost functions $c(T,v,\pi,\tau_i)$, and we explicitly explain when we require some extra assumptions on any of them. To simplify the notation, we omit the reference to $\tau_i$, because it is fixed. Numerical simulations in section \ref{Scn:ResultsPaperAnt6} utilize a particular assignment procedure (by \cite{alonso-mora_-demand_2017}) that is explained in that same section.

\subsection{Assignment introducing rewards} \label{scn:rewards}
 The first assignment method we propose consists in modifying the cost functions of feasible assignments to favor those that move the vehicles towards high-demand zones. Without any anticipatory technique, the system does not account for where the vehicle will be situated when new requests emerge, so we aim to face this issue by affecting the optimization procedure (Eq. \ref{Eq:assignment}) by reducing the costs of those routes and assignments that instruct the vehicles to move toward more convenient locations for the future.
 
Recall that $c(v,T,\pi)$ is the original cost (without anticipatory methods) of inserting trip $T$ into vehicle $v$ if the updated route including $T$ is $\pi$ (that is required to serve all requests in $T$ and all the previous requests being served by $v$). The route before inserting the new trip is  $\pi_v$. The anticipatory routing and assignment are achieved by modifying this cost function, adding a \emph{reward} $R$, which is a (negative) additive term:

\begin{equation}
\label{Eq:C_A}
  c_A(v,T,\pi)=c(v,T,\pi)-\Theta R(v,T,\pi)  
\end{equation}

The impact of Eq. (\ref{Eq:C_A}) on the system can be twofold: on the one hand, if $v$ is assigned to serve $T$, there might be more than one feasible route that fulfills all the constraints $\mathcal{C}$, so usually (and we assume this is the case for the analyses that follow) the route is chosen minimizing the cost function; thus, different routes might be selected when using $c_A$ instead of $c$. On the other hand, the decision of which vehicles assign to which trips is taken minimizing the sum of the costs of the selected assignments (Eq. \ref{Eq:assignment}), a procedure that yields different results when $c_A$ is used instead of $c$. 

Parameter $\Theta$ in Eq. (\ref{Eq:C_A}) is a tuning parameter that controls how much weight is given to the reward. Low values of $\Theta$ would have almost no impact. When $\Theta$ increases, some vehicles will take routes that are not the shortest ones and induce higher waiting times and detours for current users, so VHT (vehicle-hours-traveled) should increase, as well as users' total travel times. However, if the method effectively locates the vehicles into better positions concerning future demand, the number of rejections will decrease. Waiting times for current passengers will increase (again, because the selection of the optimal route is modified by Eq. \ref{Eq:C_A}), but might decrease for future ones, making the global effect uncertain. If $\Theta$ is too high, the system might get degraded in all its indicators, as the primary purpose of the system will be getting better prepared for the future, without looking at current users and inducing very long detours.

We provide the pseudo-code description of the method in Algorithm \ref{Alg:Rewards}. Recall that we are solving a single stage of the overall assignment problem (a specific $\tau_i$), meaning that the set $\mathcal{R}_e$ contains those requests that are to be assigned at that specific time. We assume some assignment procedure $\mathcal{P}$ is going to be modified by the rewards, and we highlight in bold the steps affected by the rewards (steps 3 and 5). Introducing rewards first affects how to determine the routes, where we admit that $\mathcal{P}$ might consider some rules (e.g., FIFO) that must be respected when optimizing (step 3), and then affects the decision of which vehicles are serving which requests (step 5). Note that even if $\mathcal{P}$ is event-based, i.e., if it assigns requests individually as soon as they appear, these modifications are still valid.

\begin{algorithm}
\caption{Introducing rewards over an assignment procedure $\mathcal{P}$.}\label{Alg:Rewards}
\begin{algorithmic}[1]
\STATE Input: The directed graph $(V,E)$, a set of requests $\mathcal{R}_e=(r_1,...,r_n)$, a set of vehicles $\mathcal{V}=(v_1,...,v_q)$, and a reward function $R$.
\STATE {Compute the feasible matches between trips $T$ and vehicles $\mathcal{V}$; \Comment{This computation is done using the original assignment procedure $\mathcal{P}$}}
\FORALL {$(T,v)$ feasible} {determine the optimal route $\pi$ \textbf{according to $c_A(v,t,\pi)=c(v,t,\pi)-\Theta R(v,t,\pi)$} and to other possible rules; \Comment{These other possible rules are given by the assignment procedure $\mathcal{P}$}}
\ENDFOR
\STATE {Determine the optimal assignment between vehicles and trips \textbf{according to $c_A$}; \Comment{How to determine this optimal assignment is given by the assignment procedure $\mathcal{P}$}}
\STATE Move the vehicles according to their updated itinerary until the assignments are re-updated;
\STATE Output: For each vehicle an updated itinerary and position, and for each request a vehicle that will serve it or a notification that it is rejected.
\end{algorithmic}
\end{algorithm}

The crucial question is how to define the reward $R$. We propose several specifications (explained in detail in section \ref{scn:rates4}); for the sake of simplicity, all of them depend on some characteristics of a particular node of the induced route $\pi$. Two questions naturally arise: which node to look at, and what to observe from that node? We study two possible answers for both questions. Let us begin with the first one:
\begin{outline}
    \1 The rewards will be a function of one specific node in $\pi$: 
    \2 The last node of $\pi$, denoted as $LN(\pi)$, or
    \2 The first node $u$ in $\pi$ verifying the following condition: the vehicle has idle capacity after visiting $u$, and it does not get full in any node afterward. This node is called ``idle node'' and is denoted $IN(v,T,\pi)$.
\end{outline}

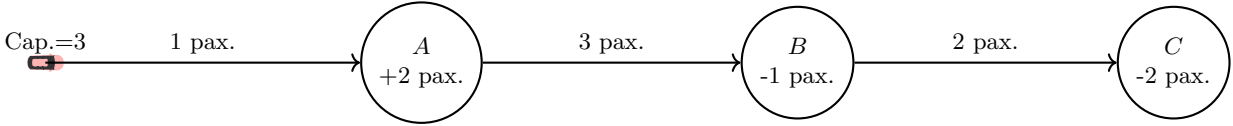
\begin{figure}[H]
    \centering
    \small
    \begin{tikzpicture}[->>=1pt,auto,node distance=1.5cm,
          thick,main node/.style={circle,draw},every text node part/.style={align=center}]
        \node[sedan top,body color=red!30,window color=black!80,minimum width=0.5cm,label=above:{Cap.=3}] at (-6,0) (V) {};
          \node[main node, ] at(-1,0) (1) {$A$ \\ +2 pax.};
           \node[main node, ] at(4,0) (2) {$B$ \\ -1 pax.};
                     
           \node[main node, ] at(9,0) (3) {$C$ \\ -2 pax.};
           

          \path [->] 
          (V) edge node {1 pax.} (1);
          \path [->] 
           (1) edge node {3 pax.} (2);
           \path [->] 
            (2) edge node {2 pax.} (3);
            
        \end{tikzpicture}
        \caption{Example of last node $LN$ and idle node $IN$. A vehicle of capacity 3 is following the route $\pi=(A,B,C)$, so its last node is $LN(\pi)=C$. The vehicle becomes full at $A$, and it stops being full at $B$, such that it is never full again. Therefore, this route's idle node is $IN(v,T,\pi)=B$.}
        \label{Fig:LNVSIN}
 \end{figure}

 An example of these definitions is provided in Figure \ref{Fig:LNVSIN}, in which a vehicle has idle capacity before its last node, so $IN$ and $LN$ do not coincide. Note that $LN(\pi)$ depends only on the route, whereas $IN(v,T,\pi)$ also depends on the vehicle and the trip. We now turn to analyze what to look at in the selected node. 

\begin{outline}
    \1 Given the node $u$ (either $u=LN(\pi)$ or $u=IN(v,T,\pi)$), the reward at time $t$ will be a function of:
    \2 Either the number of requests that departs there $Gen(u,t)$ (a \emph{generation rate}, as in \cite{vosooghi2019shared,lioris2016dynamic}), or
    \2 The number of requests that are rejected there $Rej(u,t)$ (a \emph{rejection rate}, as in \cite{alonso-mora_-demand_2017}).
\end{outline}

There might be several ways to define the generation and rejection rate of a particular node. We propose three methods, explained in section \ref{scn:rates4}. As discussed in \ref{scn:statementanticipatory}, none of these definitions shall require any knowledge about the future. We do consider a fourth method as a benchmark that is based on historical requests, which does not imply assumptions regarding the future but require exogenous data.

\vspace{5 mm}
\textbf{Analysis and discussion:} Some characteristics, virtues, and drawbacks of this method can be identified, regardless of the assignment method and of the specific scenario in which it is applied.

Parameter $\Theta$ controls the impact of the method. Note that as routes and assignments are discrete decisions, there are specific values of $\Theta$ in which the system is modified. That is, there exists $\Theta_1, \Theta_2, ..., \Theta_Q$ such that when $\Theta$ reaches each of them, some decision changes: either a vehicle changes the order in which to serve its assigned requests, or some assignment between vehicles and users is modified. Therefore, selecting $\Theta$ means deciding how many of these changes are introduced.

Let us analyze the advantages of this method. One of the main difficulties of optimizing ridepooling systems is related to the computational time required to decide the assignments: The number of feasible trips can be huge, and each trip-vehicle combination requires some computation as different routes can be followed. The rewards work over exactly the same set of feasible solutions (in other words, the set $\mathcal{A}$ from Eq. \ref{Eq:assignment} remains the same, but the cost of each $A \in \mathcal{A}$ is adapted), so if an assignment procedure has been shown to work properly in regard to the computational burden, this will continue to be true when this anticipatory method is included.

Another virtue of this method is its predictability. It affects the system in a clear-cut way, pushing some vehicles to move more often towards the most demanded zones (or with the highest rejection rates) instead of going elsewhere. Therefore, we can know in which scenarios we should expect this technique to perform well: when vehicles can choose whether to go to a high-demand zone or to a low-demand one. Note that in the extremely unbalanced scenario depicted in Figure \ref{Img:CircularCity}, vehicles have no choice but going to the unique destination, so that rewards would play no role there. However, if requests are emerging from different places within a city (which is the usual real-life case), vehicles do have a choice. 

This technique does not increase at all the computational time because it affects the system only at an ``individual level'', i.e., at each combination of vehicle and trip, which is also a disadvantage, as no global analysis of the system performance is included. In particular, all the rewards point towards the same directions, and there is no explicit control of how many vehicles are indeed directed towards each zone, so that a new imbalance might be induced. The value of $\Theta$ plays a key role here: as the rewards are included in the cost function, they will affect the behavior of the vehicles only if the reward is large enough to compensate the difference in the original costs between the routes or assignments being compared. Therefore, the larger the $\Theta$, the more number of vehicles affected. We require tuning this parameter to find intermediate values that affect the system, but not all the vehicles. 

\subsection{Assignment inserting future artificial requests} \label{scn:artif}
The method explained above has the virtue of affecting the system's decisions at every level (routing and assignments). However, it lacks a global insight into the system's performance: all the rewards are evaluated equally if evaluated in the same nodes, so all the vehicles are pointed towards the same zones. There is no mechanism to achieve a balance at a whole-network level. With this in mind, we implement a method based on \cite{alonso-mora_predictive_2017} but able to work with different generation rates (other than requiring historical data) and assignment procedures (other than the one from \cite{alonso-mora_-demand_2017}). 

 As the introducing rewards method, this method can also be used with different assignment procedures, as long as they can handle requests with future request times, and that the reject penalty from Eq. (\ref{Eq:assignment}) is a fixed value per rejected request, i.e., that the cost of rejecting a set $Q$ of requests is defined as some \textit{rejection penalty} $p_{KO}$ times the size of $Q$.

Recall that we are deciding how to assign the vehicles to a set of requests during a single stage of the whole operation. At a high-level description, this anticipatory method consists in adding to that set of requests some future artificial ones, whose origins are located in high-demand zones, so that some vehicles might get assigned to them and move towards their origins:
\begin{outline}
    \1 First, define a generation rate per node. The same generation rates used to define the rewards will be considered, which are explained in section \ref{scn:rates4}.
    \1 Second, generate $m$ artificial random requests whose:
    \2 Origins are selected randomly, following a distribution given by the generation rates.
    \2 Destinations are selected such that the length of the artificial requests is similar to the average length of the real ones. This is done to yield operator's costs that are close to the ones of the real requests. By this means, the operator's costs do not play a too determinant role when deciding whether to serve the artificial requests.
    \2 Times at which they emerge are $\tau_i+k\cdot\phi$ with $k=1 ,...,m $, where $\tau_i$ is the current time and $\phi $ is some parameter (that has time units). Artificial requests are equidistant in time to prevent them from being too close, which could make them either impossible to group (if they are close in space), or feasibly to be matched with any vehicle (in the opposite case). Therefore, $\phi$ should be large enough so that the artificial requests are not too close to each other, but not too large so that some future requests could be matched with the current ones.
    \1 The assignments are decided following the same original assignment procedure, considering both the real and the artificial requests. Artificial requests, however, present a lower rejection penalty $p_{KO}'=\Gamma \cdot p_{KO}$, where $p_{KO}$ is the rejection penalty for a real request and $\Gamma \in (0,1)$.
    \1 The artificial requests are erased after deciding the assignment and updating vehicles positions. That is, they are not kept in vehicles' lists. Therefore, they are never served, and they do not affect subsequent assignments. They only impact the immediate routes followed by the vehicles.
\end{outline}

We provide a pseudo-code description of the method in Algorithm \ref{Alg:Artificial}. The artificial requests are created in step 2 and participate in all the following steps of the assignment procedure.

\begin{algorithm}
\caption{Including artificial requests over an assignment procedure $\mathcal{P}$.}\label{Alg:Artificial}
\begin{algorithmic}[1]
\STATE Input: The directed graph $(V,E)$, a set of requests $\mathcal{R}_e=(r_1,...,r_n)$, a set of vehicles $\mathcal{V}=(v_1,...,v_q)$, and a generation rate $Gen$.
\STATE Compute $m$ artificial requests $\overline{\mathcal{R}}=\{\overline{r_1},...,\overline{r_m}\}$ according to $Gen$;
\STATE $\mathcal{R}\leftarrow \mathcal{R}_e \cup \overline{\mathcal{R}}$;
\STATE Assign requests in $\mathcal{R}$ to vehicles in $\mathcal{V}$ according to the assignment procedure $\mathcal{P}$;
\STATE Move the vehicles according to their updated itinerary until the assignments are re-updated;
\FORALL{$v \in \mathcal{V}$} Remove every artificial request (if any) in the itinerary of $v$;
\ENDFOR
\STATE Output: For each vehicle an updated itinerary and position, and for each real request a vehicle that will serve it or a notification that it is rejected.
\end{algorithmic}
\end{algorithm}

\textbf{Analysis and discussion: } First, let us understand how the system compares the costs of future requests with those of real requests. Users' costs of a future request are usually low because vehicles can arrive before the future request yielding zero waiting times. On the other hand, the destinations of futures requests are defined so that the operator's costs are similar for real and future requests. Therefore, parameter $\Gamma$ plays a key role. If $\Gamma$ is close to $1$, i.e., if the rejection penalty is almost the same for future and real requests, then artificial requests are less costly (due to the users' costs) than the real ones, and the system will always prioritize them. In the opposite case, if $\Gamma \approx 0$ then future requests will never be selected because rejecting them is almost costless. Therefore, it is crucial to select a balanced value for this parameter, so that the system selects only those artificial requests that could be served very efficiently, i.e., when moving the vehicles towards their origins is not too costly.

A virtue of this method is that it considers the set of vehicles as a whole. Indeed, future requests are assigned to the vehicles together with the current ones, so no more than one vehicle will be directed to each artificial request. Moreover, an artificial request might be rejected if serving it is too costly, i.e., if the system needs to make a considerable effort to move some vehicle towards its origin. Therefore, this mechanism is balanced at a whole-network level.

Although the method directly affects the algorithm only when deciding the assignments, routes are altered as well, and differently than when introducing rewards. To see this, consider the following example: a vehicle is on its way to pick up a passenger $p$, but the new assignment sends it to pick-up a future request $r$ first. Although the vehicle is not necessarily arriving at the origin of $r$, its route is modified until the system updates its assignments. Therefore, when this method is applied, the vehicle does not always follow the shortest paths between consecutive stops. Although this induced detour might be seen as a drawback, it is actually something pursued by the anticipatory method, as the new route will likely cross some high-demand areas that increase the chance of receiving a new passenger while driving there.

However, there are also drawbacks. This technique increases the computational load, as new requests imply new feasible trips. Moreover, as the emerging time of the artificial requests is in the future, a large portion of the vehicles can get to their origins before the respective request times; therefore, most feasible trips containing only real requests can be combined with many subsets of future requests, leading to an increment in the number of trips that might be exponential. In many ridepooling algorithms (like the one proposed by \cite{alonso-mora_-demand_2017}, which we use here in the simulations), computing the feasible vehicle-trips matching is the heaviest burden, which might become much slower with future requests.

Moreover, the impact of the artificial requests on the system is hard to predict. Three types of effects can be identified: first, some vehicles that could have become idle without this method will move towards the artificial origins, which is similar to the rebalancing steps that are usually present in these models (see section \ref{scn:related}); second, some vehicles will add future requests to their non-empty lists of requests they are serving, which will modify the route in which these real requests are visited; and third, some vehicles might have to decide if serving a future request or a real one. The first effect is not too relevant, as a rebalancer could replace it; the second effect is the one we pursue; the third effect might be troublesome because if a vehicle prioritizes serving the future request, the technique will induce some rejections that could be saved. Again, selecting a $\Gamma$ that allows for the second effect and not for the third one is ideal, but it might not be possible.

\section{Different definitions for generation and rejection rates} \label{scn:rates4}
There can be several ways to define the generation and rejection rates for a specific node. Here we focus on definitions that can be computed utilizing only the information that is generated by the system while operating, so that we do not require any exogenous information. Namely, we will look at the origins of the requests that have recently emerged (for the generation rates), or that have been recently rejected (for the rejection rates).

To be concrete, we consider three definitions: a basic one ($Gen_B, Rej_B$), a smoothed one ($Gen_S, Rej_S$), and one based on particle filters ($Gen_{PF}, Rej_{PF}$). Moreover, we also consider an additional way to compute the generation rates that is based on historical data ($Gen_H$), so that we can evaluate whether using recent information is more appropriate; we cannot use historical data to compute a rejection rate, as rejections depend not only on the exogenous requests but also on the operation of the system. The first two methods are calculated for each node, whereas the other two need to first cluster the nodes in the graph into zones. The temporal evolution of the system plays a role in the definitions that follow, so it is worth including the explicit reference to the decision times $\tau_i$. We now explain each rate in detail:

\subsection{Basic rates}
The simplest way to define the generation (rejection) rate of a node $u$ is to look at the number of requests that have just been generated (rejected) at $u$. Note that the generation rate depends only on the set of requests, and the rejection rate also depends on how they are assigned. Denoting $Rej(\tau_i,u)$ as the number of requests emerging from $u$ that were rejected by the system at the corresponding assignment, the basic rates are defined by:

\begin{equation} \label{EqBasic}
    Gen_B(u,\tau_i)=|\{r \in \mathcal{R}_{e,\tau_i}: o_r=u \}|, Rej_B(u,\tau_i)=Rej(\tau_{i-1},u)
\end{equation}

For $i=1$, the rejection rates are defined as zero everywhere.

Eq. \ref{EqBasic} considers, for each node $u$, the number of requests that emerged since the last assignment, and the number of requests rejected in the last assignment, respectively. The intuition is straightforward: a node having a large generation rate means that many users currently require a vehicle there, so it is worth directing part of the fleet in that direction. On the other hand, a large rejection rate represents a lack of vehicles in the last assignment. 

Despite being an intuitive method, it might be somewhat unstable: as the $\tau_i$ cannot be too widely spaced in time, the chance that a node has no requests is high, and the randomness might play a too relevant role.

\subsection{Smooth rates}
The aim of anticipatory routing is that vehicles remain closer to where demand is expected. From that point of view, the rates of a node could also consider the information of its neighboring nodes: for instance, it might be better for a vehicle to be in a node that does not generate requests if all its neighbors do. With this in mind, the $Gen_S$ and $Rej_S$ rates are defined considering also the requests that depart from close nodes:
\begin{equation}
\label{Eq:Smooth}
    Gen_S(u,\tau_i) = \sum_{w \in V} \frac{Gen_B(w,\tau_i)}{\psi + t_V(u,w)}, Rej_S(u,\tau_i) = \sum_{w \in V} \frac{Rej_B(w,\tau_i)}{\psi + t_V(u,w)}.
\end{equation}
Where $t_V(x,y)$ is the time-length of the fastest path between $x$ and $y$, and $\psi$ is a tuning parameter (the higher this parameter, the more uniform the resulting rates). This method is called ``smooth rates'' because rates become more stable in space. Note that all nodes are included in Eq. (\ref{Eq:Smooth}), but distant nodes do not affect much.

\subsection{Particle filters} \label{subsection:PF}
This method applies the ideas from \cite{wallar2018vehicle} (based on the particle filter methods proposed by \cite{arulampalam2002tutorial}) to calculate the generation/rejection rates of a zone, although \cite{wallar2018vehicle} used it for rebalancing purpose. It requires first dividing the nodes into clusters $C_1,...,C_M$, which are obtained by minimizing the number of required ``centers'', such that each node in the network can be reached from at least one center in a time lower than a parameter $t_M$. This problem is solved through an ILP that is fully described in \cite{wallar2018vehicle}, and each node is then assigned to its closest center. All the nodes assign to the same center comprise a zone. Note that by this method all the zones have a similar area. This is crucial to have rates that are comparable (otherwise, larger zones could present higher absolute rates but having a lower density of requests, so that it would be unclear whether to prioritize them).

Once the $M$ centers and zones have been obtained, auxiliary variables (that represent a proxy for the rates) $\lambda_{z\ell}$ are defined for each zone $z$ and for $\ell=1,...,\eta$, where $\eta$ is a large number of Montecarlo simulations. Each $\lambda_{z\ell}$ has a weight $w_{z\ell}$. These values will be updated at each $\tau_i$, and the idea is that the rates at each zone are calculated as the weighted averages of the $\lambda_{z\ell}$. To simplify the notation, we omit the reference to $\tau_i$ when referring to variables $z$ and $w$.  Parameters $\lambda_{z\ell}$ are initialized randomly, with $w_{z\ell}=\frac{1}{\eta}$. Denoting $G(z,\tau_i)=\sum_{u \in z} Gen_B(u,\tau_i)$ the generation rate of each zone, let us explain how rates are updated (the procedure is analogous for the rejection rates):
\begin{enumerate}
    \item  For each zone $z$, $\eta$ samples are selected (with replacement) from $\lambda_{z\ell}$, with probabilities given by the weights $w_{z\ell}$. These samples are denoted $\overline{\lambda_{z\ell}}$.
    \item Variables $\overline{\lambda_{z\ell}}$ are perturbed\footnote{An almost-zero lower bound is required because negative rates are senseless. In our experiments, this bound was never reached}: $\lambda_{z\ell}=\overline{\lambda_{z\ell}}+\mathcal{\eta}(0,\sigma^2)$, with $\sigma^2$ a ``volatility parameter''.
    \item Weights are updated according to the observed generation rate of the zone, as the probability that a Poisson process of parameter $\lambda_{z\ell}$ reaches the generation rate of the zone: $\overline{w_{z\ell}}=e^{-\lambda_{z\ell}}\frac{\lambda_{z\ell}^{G(z,\tau_i)}}{G(z,\tau_i)!}$. By these means, those $\lambda_{z\ell}$ that are closer to the observed rates get a higher weight, which affects the calculated generation rate (step 5) and the next update (in $\tau_{i+1}$) for $\lambda_{z\ell}$ (step 1).
    \item Weights are normalized: $w_{z\ell}=\frac{\overline{w_{z\ell}}}{\sum_{j=1}^N \overline{w_{zj}}}$
    \item Generation rates for each zone are calculated as the weighted averages: $Gen_{PF}(z,\tau_i)=\sum_{\ell=1}^\eta w_{z\ell} \lambda_{z\ell}$. Each node $u \in z$ has the same generation rate $Gen_{PF}(u,\tau_i)=Gen_{PF}(z,\tau_i)$.
\end{enumerate}
Using zones instead of single nodes might make the system more robust as it does not depend on events on single nodes, but changes in the boundaries of a zone might be troublesome. The particle filter method has the virtue of providing some temporal stability, as the parameters $\lambda_{z\ell}$ are updated at each $\tau_i$ from their previous values (see step 3 above). However, the stability might be a problem when big changes take place (for instance, when passing from the peak to the off-peak period).

\subsection{An additional definition and comparison of the rates}
As explained above, we only use information that is endogenously generated by the system when assigning. The only exception is the rates we explain now, which are based on exogenous historical data so that they might be used as a benchmark to analyze how useful current requests are to approximate what will happen in the near future. Note that using historical data does not violate the principle of not requiring any assumptions regarding future requests (e.g., in the form of a probability distribution).

This method adapts the ideas from \cite{alonso-mora_predictive_2017}. They also divide the nodes into clusters, so we keep here the technique used for the particle filter method (previous subsection). Then, they estimate the number of requests emerging from a zone using a historical dataset. Which dataset to use is not a trivial issue, as transport demand can be heavily affected by weather, traffic events, recent transport-related changes -such as new transit/metro lines or new highways- or urban projects, among others \cite{bocker2013impact,zhou2013passenger,xue2015short,pereira2014role,liu2021weather}, which makes data-based demand prediction a quite complex challenge, beyond the scope of this paper. Nevertheless, we do take this discussion into account: instead of considering a whole year of data (as \cite{alonso-mora_predictive_2017}), we use only some weekdays in the past, which are expected to have more similar weather (same season) and fewer differences in the private and public transport networks. Denoting $\mathcal{D}$ the set of days in the dataset, and $G(u,d,t_1,t_2)$ the number of requests emerging from node $u$ during $(t_1 ,t_2)$ on day $d$:

\begin{equation}
    Gen_H(z,\tau_i)=\sum_{u \in z}\sum_{d \in \mathcal{D}} \frac{G(u,d,\tau_{i-1},\tau_i)}{|\mathcal{D}|}
\end{equation}
With all the zones beginning with a nil generation rate. And
\begin{equation}
    Gen_H(u,\tau_i) = Gen_H(z,\tau_i) \forall u \in z
\end{equation}

We only use this method for generation rates, as there is no such thing as ``historical rejection rates''. The advantages of this method relate to the potential predictive power that historical data might provide, which can be particularly relevant when the demand faces sudden changes that will not be anticipated by methods that rely on the current and past states of the system. Therefore, it can be used to compare the results obtained by the other methods. On the other hand, historical data is not always available, and using it effectively for predicting the demand is not an easy task.

\begin{figure}[H]
    \centering
\qquad
\subfloat{{\includegraphics[width=3cm]{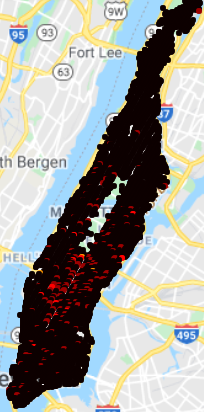} }}
   \qquad
\subfloat{{\includegraphics[width=3cm]{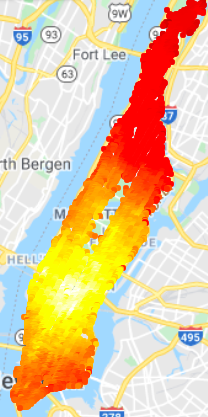} }}
\qquad
\subfloat{{\includegraphics[width=3cm]{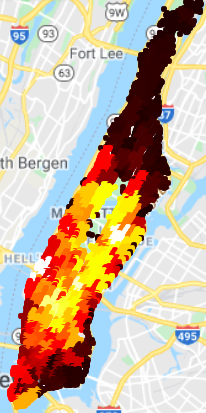}} }
\qquad
\subfloat{{\includegraphics[width=3cm]{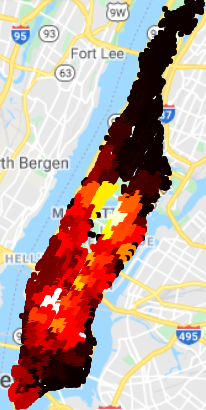}} }
\caption{The generation rate of each node according, from left to right, to $Gen_B, Gen_S, Gen_{PF} \text{, and } Gen_H$ in Manhattan, at 12:50 on 15/01/2013, considering requests emerged in the previous minute. The brighter the color, the higher the generation rate.}
\label{Img:GenRates}
\end{figure}

A graphical comparison of the four methods based on generation rates is offered in Figure \ref{Img:GenRates}, in which each node's color represents its generation rate at a particular time (the brighter the color, the higher the rate). Although all the methods present higher rates at the center of the network, the differences among them are highlighted as follows: in the basic method, most nodes present nil generation rate (black nodes), with a few nodes concentrating all the positive rates; the smooth method builds a much more continuous pattern, which yields positive values even in the extreme sectors of the network; the last two methods are based on zones, with the particle filter achieving a higher concentration at the center (brighter colors). The rejection rates cannot be shown at this point, as they depend on the specific conditions of the ridepooling system and not only on the demand itself. 

The apparent differences among the resulting rates suggest that the effectiveness of the anticipatory techniques shall depend greatly on which rates are being used. Recall that we aim for balanced values for the parameters $\Theta$ and $\Gamma$ that control the weight of the anticipatory techniques. The larger the differences among the generation rates per node, the more significant the differences introduced by the anticipatory techniques when deciding the routes and assignments. In other words, while $\Theta$ and $\Gamma$ refer to the overall weight between current requests and anticipatory techniques, the intensity of a method refers to how distributed (more or less concentrated) this weights are in space. Therefore, having such varying rates provides yet another source of flexibility (together with the parameters $\Theta$ and $\Gamma$) to control the anticipatory methods, which might enable better results. 

\section{Numerical simulations}
 \label{Scn:ResultsPaperAnt6}
In this section, we run simulations using over a real-life case using the assignment procedure by \cite{alonso-mora_-demand_2017}. We first explain such a procedure (section \ref{Section:assignment5}), and then describe the real-life case (section \ref{sec:reallife}) to present detailed results and sensibility analyses (sections \ref{sec:globalperformance}-\ref{Sctn:sensibility})).
\subsection{The assignment procedure} \label{Section:assignment5}
The anticipatory ideas we propose in this paper are tested on top of the assignment procedure studied by \cite{alonso-mora_-demand_2017}, which assigns passengers to vehicles. This procedure is based on deciding each $\delta $ (here we use $\delta $ = 1 minute) how to assign the requests that have emerged during that lapse of time: how to group them, which vehicle (that might be serving some previous passengers) assign to each group, and in which order, i.e., $\Psi=\{k \cdot \delta: k=1,...,\ceil{PO/\delta}\}$. The set of constraints $\mathcal{C}$ requires that the capacity of the vehicles is never exceeded, and deals with the quality of service, imposing that no served requests can face a waiting time large than some upper bound $\Omega_w$, nor a total delay larger than $\Omega_d$.  We now explain how the assignments are decided in each iteration at time $\tau_i$.

\begin{enumerate}
    \item First, for each request $r \in \mathcal{R}_{e,\tau_i}$ and vehicle $v$ it is analyzed if it is feasible that $v$ serves $r$ without violating the constraints $\mathcal{C}$. A heuristic might be implemented after this step to reduce the number of feasible requests (and thus, the computational time): for each request $r$, consider the set of all feasible to serve $r$, and discard the most costly ones.
    \item Assuming that the feasible assignments between vehicles and trips of size $j$ are known (the previous bullet point explains the case $j=1$), study the feasible matches between vehicles and trips of size $j+1$ based on the following fact: if an assignment between a trip and a vehicle is feasible, then the assignment between the same vehicle and all the subsets of that trip must be feasible as well. When a feasible link between a vehicle and a trip is found, it includes how to update the vehicle's route, which is done by minimizing the total additional costs (see Eq. \ref{eq:CUCO} below for details), which can be done using an exhaustive search or some insertion heuristic.
    \item Build an ILP that decides which of the feasible trip-vehicle assignments are taking place, ensuring that each request is either assigned to a single vehicle or rejected, and that each vehicle is assigned to no more than one trip. The objective function includes the costs of the chosen assignments, plus a penalty $p_{KO}$ for each rejected request.
    \item A rebalancing step moves the idle vehicles, i.e., those that had no passengers before deciding the assignment and received none in step 3. They are moved towards the origins of the rejected requests through another ILP, minimizing the total distance driven by these vehicles and without sharing (i.e., no more than one rejected request can be assigned to each idle vehicle). Note that these vehicles are not actually meant to serve those rejected requests, so the feasibility constraints regarding waiting times, delay, and capacity of the vehicles are not included in this ILP.
\end{enumerate}

Some comments are noteworthy:
\begin{itemize}
    \item In \cite{alonso-mora_-demand_2017}, when the assignment is decided, those requests that are not picked up before $\delta $ are kept for the next iteration, which allows the system to reassign. That is to say, these requests might be served by a different vehicle or can even become rejected, if doing so increases the total efficiency of the system. We introduce a slight modification in this paper: the first request that will be picked up (i.e., the next pick-up in the vehicle's list) is not reassigned in the next iteration. This change is needed because when routing is modified through the anticipatory techniques, the time required to arrive at the first pick-up might increase because the vehicle does not necessarily follow shortest paths; therefore, the number of accumulated requests for reassigning might increases, which has an exponential impact on the computational burden. Moreover, experiments show that this change reduces the number of rejections in the system.
    \item If a reassignment of an individual request takes place, \cite{alonso-mora_-demand_2017}  forbids an increase in the achieved waiting time (compared to the one of the previous assignment), while in \cite{fielbaum2021demand} they forbid rejections for a subset of the requests (those that are required to walk, an option that is not considered here). In this paper, requests that are being reassigned face the same constraints as the new requests and can be rejected.
    \item Once a request becomes rejected, it is removed from the system, instead of kept for reassigning as done in \cite{alonso-mora_-demand_2017}.
\end{itemize}

We use the following cost function when assigning trip $T$ to vehicle $v$, if the route of the vehicle is updated from $\pi_{v,\tau_i}$ to $\pi$ (prior to introducing rewards):

\begin{equation*}
c(v,T,\pi)=\sum_{r \in T}{\left[p_w t_w(r,\pi)+p_v D(r,\pi)\right]} + \sum_{r_0 \in Req_{v,\tau_i}}{\left[p_w \Delta t_w(r_0,\pi,\pi_{v,\tau_i})+ p_v \Delta D(r_0,\pi,\pi_{v,\tau_i})\right]}
\end{equation*}

\begin{equation}
\label{eq:CUCO}
    +p_O  \left( L(\pi)-L(\pi_{v,\tau_i}) \right)
\end{equation}

Where the first line in Eq. (\ref{eq:CUCO}) represents users' costs: the first term refers to the waiting time $t_w$ and the detour $D$ faced by the requests $r$ in the new trip, and the second term to the extra waiting time $\Delta t_w$ and extra detour $\Delta D$ imposed to the requests $r_0$ that were already being served by the vehicle $v$, due to the changes induced to its route. Parameters $p_w$ and $p_v$ are the costs of one time-unit waiting and over the vehicle, respectively. The second line in Eq. (\ref{eq:CUCO}) deals with the operator's costs, which are proportional to the increase in the length of the route. 

A detailed explanation of the assignment method per iteration, already affected by the anticipatory techniques, is shown in the pseudo-code of Algorithm \ref{Alg:AssignmentRewards} that introduces rewards, where we say that $c(v,T,\pi)=\infty$ if the assignment is unfeasible. The algorithm including future requests is the same, but considering $c$ instead of $c_A$, and adding the future requests to $R$ at the beginning of the algorithm.

\begin{algorithm}
\caption{The assignment algorithm with rewards utilized in the numerical simulations.}\label{Alg:AssignmentRewards}
\begin{algorithmic}[1]
\STATE {Input: The directed graph $(V,E)$ and a set of requests $\mathcal{R}_e=(r_1,...,r_n)$, a set of vehicles $\mathcal{V}=(v_1,...,v_q)$. \Comment{The set of requests contains those that have just emerged, and those that might be reassigned}}
\STATE {$T=[\:], C=[\:]$;\Comment{$T$ will contain the feasible trip-vehicle assignments, $C$ will contain the respective costs}}
\\
\FORALL{$r \in \mathcal{R}_e, v \in \mathcal{V}$} 
\IF {$\exists \text{ route } \pi $ such that $c(v,r,\pi)<\infty$}
\STATE{Find $\pi'$ that minimizes $c_A(v,r,\pi)$; \Comment{$c_A$ is the cost function modified with the rewards. This route can be found through an exhaustive search or using an insertion heuristic.}}
\STATE{$T\leftarrow [T,(r,v)], C \leftarrow [C, c_A(v,r,\pi']$};
\ENDIF
\ENDFOR
\\
\FORALL{$r \in \mathcal{R}_e$}
\STATE{Remove from $T \text{ and } C$ a portion of the assignments involving $r$, those with the highest costs in $C$; \Comment{This is a heuristic step, which is optional. The more assigments that are removed, the higher the impact of the heuristic.}}
\ENDFOR
\\
\FORALL {$v \in \mathcal{V}$}
\STATE{Define $R_v=\{r \in \mathcal{R}_e: (r,v) \in T\}$; \Comment{The requests that can be served by $v$}}
\FORALL{$S \subseteq R_v$}
\IF {$\exists \text{ route } \pi $ such that $c(v,S,\pi)<\infty$}
\STATE{Find $\pi'$ that minimizes $c_A(v,S,\pi)$; }
\STATE{$T\leftarrow [T,(S,v)], C \leftarrow [C, c_A(v,S,\pi')]$};
\ENDIF
\ENDFOR
\ENDFOR
\\
\STATE{Select a subset $T'$ from $T$, such that each vehicle and request are in no more than one element in $T'$, minimizing the total cost of $T'$ plus the penalty for each non-assigned requests; \Comment{$T'$} contains the assignments that are taking place. The optimization problem is solved as an ILP.}
\STATE Output: $T'$

\end{algorithmic}
\end{algorithm}

It is worth recalling that the anticipatory methods we propose here do not require this specific assignment procedure to work. Different assignment methods and different cost functions may also incorporate the anticipatory techniques explained above.

\subsection{The real-life study case} \label{sec:reallife}
The proposed methods are tested over a publicly available dataset of real trips performed by taxis in Manhattan, New York, that started between 1-2 p.m. on January 15th, 2013. The total number of requests is 7,748, while 4,091 nodes and 9,452 edges form the city network. The numeric value of the chosen parameters is shown in the Appendix. Some conditions are modified to analyze the robustness of the system in section \ref{Sctn:sensibility}.

A fleet of 1,000 vehicles of capacity 3 was considered. This is a small fleet, unable to serve all the demand. Having a significant rejection rate enables us to analyze the impact of the methods over rejections in a crisp way, as well as how anticipatory techniques affect the trade-off between the number of served requests and the quality of service for those who are served. In section \ref{Sctn:sensibility}, we show that the main analyses and conclusions we obtain for this fleet remain valid when a larger fleet is used.

Let us first analyze the results obtained when only the methods that provide rewards at the routing stage are included (sections \ref{scn:finaloridle} and \ref{scn:rewardscomparingrates}). We then study the effects of the artificial requests (section \ref{scn:resultsartficialrequests}), we compare the methods (section \ref{scn:coupling}), we study in detail the operational effects of the anticipatory techniques (section \ref{scn:whatishapp}), and we finally provide a sensibility analysis (section \ref{Sctn:sensibility}).

\subsection{Global performance of the anticipatory methods} \label{sec:globalperformance}
\subsubsection{Assignment introducing rewards: Final node or first idle node} \label{scn:finaloridle}
\begin{figure}[H]
    \centering
\qquad
\subfloat{{\includegraphics[width=4.2cm]{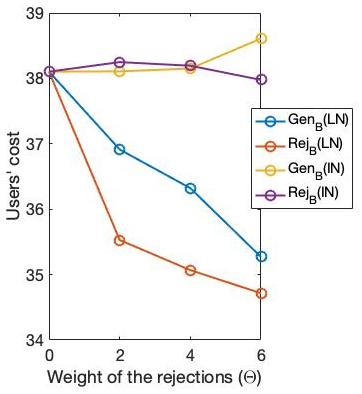}} }
   \qquad
\subfloat{{\includegraphics[width=4.2cm]{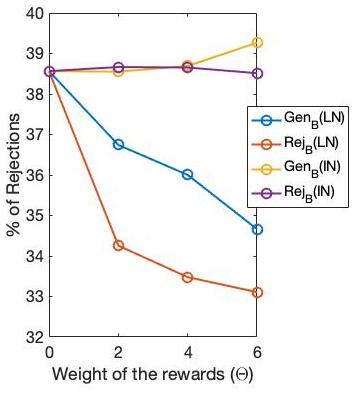}} }
\qquad
\subfloat{{\includegraphics[width=4.3cm]{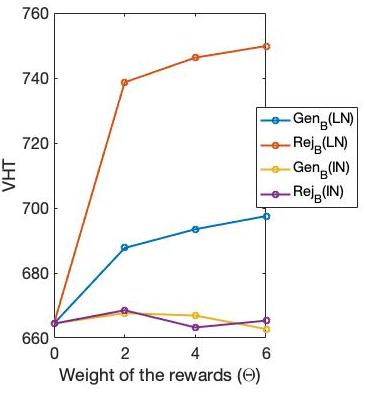}} }
\caption{Average users' costs (left), percentage of rejected requests (center), and vehicle-hours-traveled (right) in the basic model introducing rewards, as a function of the parameter $\Theta$, depending on whether the last node (LN) of the route is considered or the first node from which the vehicle always has an idle capacity (IN). $\Theta=0$ represents no anticipatory methods.}
\label{Img:IdleVSFinal}
\end{figure}

Recall that, as explained in section \ref{scn:rewards}, rewards depend on the generation rate of one of the nodes in the route that is being analyzed. Figure \ref{Img:IdleVSFinal} compares the results obtained when using the final node of the route (``Last node'' in the images) versus using the first detention of the route from which there is always idle capacity (``Idle node'' in Figure \ref{Img:IdleVSFinal}), when using the basic rates, and considering three indicators of the quality of the system: Average users' costs -an adimensional value calculated as the weighted average of waiting times, total delays and number of rejections, according to the same parameters used in the optimization, including the rejection penalty-, percentage of requests being rejected and VHT.

Results in Figure \ref{Img:IdleVSFinal} are questionless. When the reward depends on the idle node of the route, the impact is mild, but when it depends on the last node, it can be quite meaningful: compared to the case with no rewards ($\Theta=0$), the number of rejections can be reduced by a bit more than 10\%, at the cost of increasing VHT in about 10\%. Users' costs also drop, but less significantly than rejections, meaning that delay and waiting times might increase. These changes will be studied in more detail in the following subsections that compare the different rates and methods. From now on, all the results are calculated with the rewards depending on the final node.

\subsubsection{Assignment introducing rewards: Comparison of the different rates} \label{scn:rewardscomparingrates}
In section \ref{scn:rates4}, four definitions for the rates were provided: basic ($B$), smooth ($S$), calculated through particle filters ($PF$), and through historical data ($H$). The first three ones can be applied to generation or rejection rates of each node (that, as just explained, is the final node of the route), whereas historical data only gives generation rates. All this together makes seven methods to define the different rates.

\begin{figure}[H]
    \centering
\qquad
\subfloat{{\includegraphics[width=4.2cm]{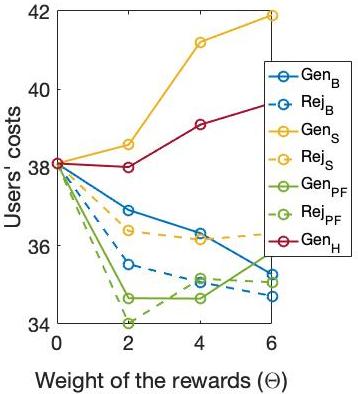}} }
   \qquad
\subfloat{{\includegraphics[width=4.3cm]{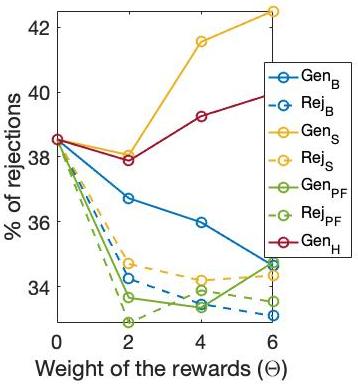}} }
\qquad
\subfloat{{\includegraphics[width=4.4cm]{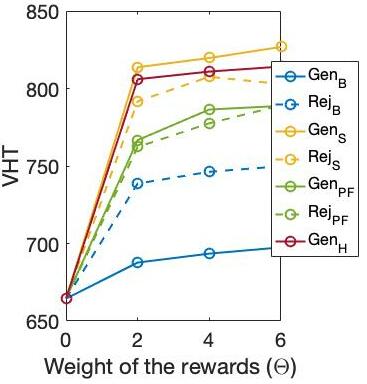}} }
\caption{Average users' costs (left), percentage of rejected requests (center), and vehicle-hours-traveled (right), as a function of the parameter $\Theta$, for each of the generation and rejection rates when introducing rewards. Solid lines represent results based generation rates, whereas dotted lines represent results based on rejection rates. $\Theta=0$ represents no anticipatory methods.}
\label{Img:7methods}
\end{figure}

The results obtained with each of these seven rates are shown in Figure \ref{Img:7methods}. The following conclusions emerge:
\begin{itemize}
    \item Some of the rates are much more effective in reducing the number of rejections and users' costs than others. Two rates highlight as the best ones: $Rej_{PF}$ and $Gen_B$. While the former achieves the minimum values for the rejection rates and the users' costs, the latter achieves slightly worse results in those measures, requiring a lower increase in VHT. It is worth saying that if $\Theta$ is further increased, $Gen_B$ does not longer improve its results.
    \item The results of the best methods show that these rewards can be very fruitful. For instance, the number of rejected requests with $Rej_{PF}$ drops from 3,025 to 2,631. Of course, whether this is good news depends on how opposing objectives are evaluated, as increasing VHT is unavoidable.
    \item The tuning parameter $\Theta$ emerges as a crucial issue for these models. Which is the best $\Theta$ depends on what rate is being used. The good news is that for all of them, the range of values of $\Theta$ that yields good results is wide, meaning that these methods are robust even if the optimal $\Theta$ is not known.
    \item Rates that are based on rejections have, in general, better results concerning both lower users' costs and VHT. The only exception is $Gen_B$, which increases VHT much less than $Rej_B$, with similar (although worse) results regarding users' costs.
    \item Regarding smoothing rates, they can improve the system if based on rejections.
    \item The rates based on historical data perform worse than the ones based on recent information. That is to say, our results highlight the potential of utilizing the information that is directly generated by the system.

\end{itemize}

\begin{figure}[H]
    \centering
\qquad
\subfloat{{\includegraphics[width=4.6cm]{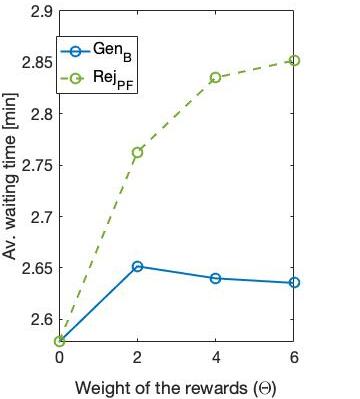}} }
   \qquad
\subfloat{{\includegraphics[width=4.6cm]{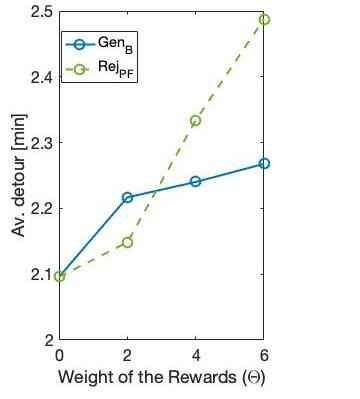}} }

\caption{Average waiting times (left) and detours (right), as a function of the parameter $\Theta$, for $Gen_B$ and $Rej_{PF}$, when introducing rewards. $\Theta=0$ represents no anticipatory methods.}
\label{Img:TWDetour}
\end{figure}

To understand the trade-off between the different user-related quality measures better, Figure \ref{Img:TWDetour} shows the average waiting time and average detour (i.e., the difference between each user's in-vehicle time and the time-length of the shortest path he/she could have followed), considering only the two rates that achieved the best results $Gen_B$ and $Rej_{PF}$. Indeed, both quality measures get worse when rewards are applied. However, recalling that $Gen_B$ achieves its minimum reject rate at $\Theta=6$ and $Rej_{PF}$ at $\Theta=2$, we can also conclude that there is some synergy between these objectives, as waiting times and detour do not increase that much for these optimal values of $\Theta$. 

As a synthesis, \textbf{the introduction of rewards reduces the percentage of rejections of the system if the right generation or rejection rates are selected, at the cost of providing worse service for the users that are transported}.

\subsubsection{Assignment inserting future artificial requests: Comparison of the different rates} \label{scn:resultsartficialrequests}
The same four generation rates $Gen_B, Gen_S, Gen_{PF}$ and $Gen_H$ were used to determine the origin of $m$ artificial requests. We take $\phi=\delta$ (1 minute). Inserting future requests makes the algorithm much slower, as they can be combined with most of the current requests without violating the constraints regarding maximum waiting times and delay, which increases the number of feasible groups. To overcome this issue, we tighten the heuristic explained in step 2 of the assignment model (section \ref{Section:assignment5}), i.e., we discard a larger amount of feasible-but-costly vehicles when analyzing trips of size one. 

\begin{figure}[H]
    \centering
\qquad
\subfloat{{\includegraphics[width=4.9cm]{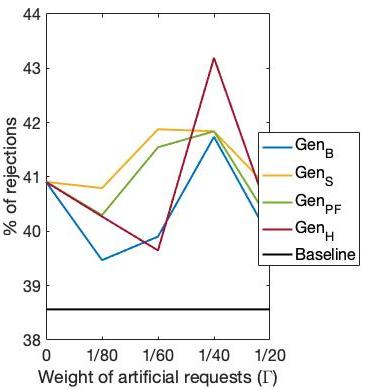}} }
   \qquad
\subfloat{{\includegraphics[width=5cm]{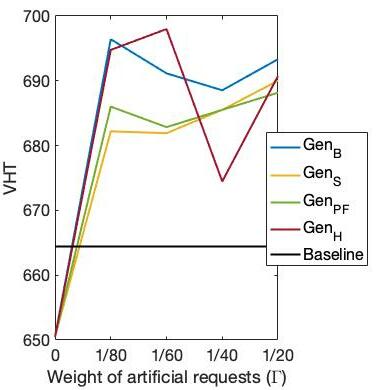}} }
   
\caption{Rejection rates (left), and vehicle-hours-traveled (right), when inserting artificial requests as a function of the parameter $\Gamma$, for each of the four generation rates. The baseline results, i.e. with no anticipatory methods and without modifying the heuristics of the assignment algorithm, are shown in black. }
\label{Img:ResultsArtificial1}
\end{figure}

\begin{figure}[H]
    \centering
\qquad
\subfloat{{\includegraphics[width=5cm]{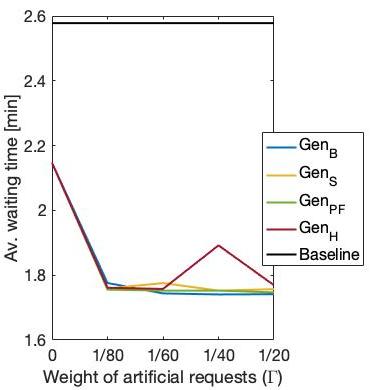}} }
   \qquad
\subfloat{{\includegraphics[width=5cm]{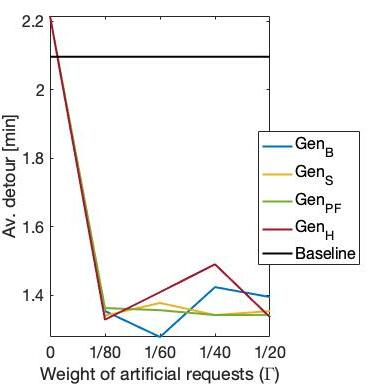}} }
   
\caption{Average users' waiting time (left), and detour (right) when inserting artificial requests, as a function of the parameter $\Gamma$, for each of the four generation rates. The baseline results, i.e. with no anticipatory methods and without modifying the heuristics of the assignment algorithm, are shown in black.}
\label{Img:ResultsArtificial2}
\end{figure}

Figures \ref{Img:ResultsArtificial1} and \ref{Img:ResultsArtificial2} show the results when inserting the artificial trips with each of the generation rates. The baseline (in black) corresponds to the results shown in the previous subsection for $\Theta=0$ (i.e., without tightening the heuristic that discards vehicles), whereas the results for $\Gamma=0$ include the change on the heuristic and are equivalent to having no artificial requests. The comparison of the results against $\Gamma=0$ shows the direct effect of introducing these artificial requests. However, the most relevant comparison is against the baseline because it is achieved if no artificial requests are added. Note that $\Gamma=0$ (i.e., when the heuristic is tightened) includes more rejections than the baseline, but better results in the other indices (much lower waiting times and detours just a bit larger), which is a natural consequence of removing the most costly vehicles for each request: each request has fewer options to be served, but the options that remain are less costly, i.e., provide a better quality of service (recall that the cost is defined as the sum of users' costs and operator's costs).

In general, all the rates are able to reduce waiting times and detours significantly. The percentage of rejections, on the other hand, is always higher than in the baseline: when compared with $\Gamma=0$, rejections are sometimes larger and sometimes fewer, but changes are always minor (these results are similar to the ones obtained by \cite{alonso-mora_predictive_2017}). 

These results can be synthesized by stating that \textbf{inserting artificial requests by itself improves the quality of service for those users that are transported, but the induced increase to the computational time requires using heuristics that might increase the number of rejections.} The interpretation is as follows: Artificial requests effectively push the vehicles towards the origins of future requests. However, when they are inserted, they compete with current requests for the same vehicles so that sometimes the system will prioritize serving the artificial ones despite their lower rejection penalty. VHT always increases. 

The method that achieves the best results is $Gen_B$: it presents the largest reduction in the rejection rate with $\Gamma=\frac{1}{80}$, and in detours for $\Gamma=\frac{1}{60}$. Reductions in waiting times are similar for all the generation rates, except for $Gen_H$ when $\Gamma=\frac{1}{40}$. Results obtained by $Gen_S$ and $Gen_{PF}$ are almost identical. The fact that $Gen_H$ might yield the worst results if $\Gamma$ is not properly selected (i.e., it is a less robust method) reinforces the conclusion that the direct use of past information can be an unfruitful idea for these transportation systems.

In all, the insertion of future requests  achieves reductions precisely where rewards fail: waiting times and detours. This happens because both methods move the vehicles towards high-demand zones, but inserting future requests might yield to rejecting some real current ones, so that the gains in efficiency translate into waiting and delay for the requests that will emerge afterward.

\subsubsection{Coupling and comparison of the methods} \label{scn:coupling}
The different results obtained by the two methods proposed in this paper, introducing rewards and inserting artificial requests, might suggest that using both simultaneously is promising, as they achieve good performance in complementary indices. However, our simulations show the opposite. As they both push the system towards the same direction, the results are much worse when used together, as rejection rates jump from less than 40\% in the baseline scenario to more than 50\%. This bad news can be explained: rewards move the vehicles towards the same zones in which the future requests are inserted. If a vehicle has to choose between waiting for a future artificial request or serving a current real one, it might prioritize the future one despite its lower rejection rate, as being in a close position makes the cost of serving it very low.

\begin{figure}[H]
    \centering
\qquad
\subfloat{{\includegraphics[width=5.25cm]{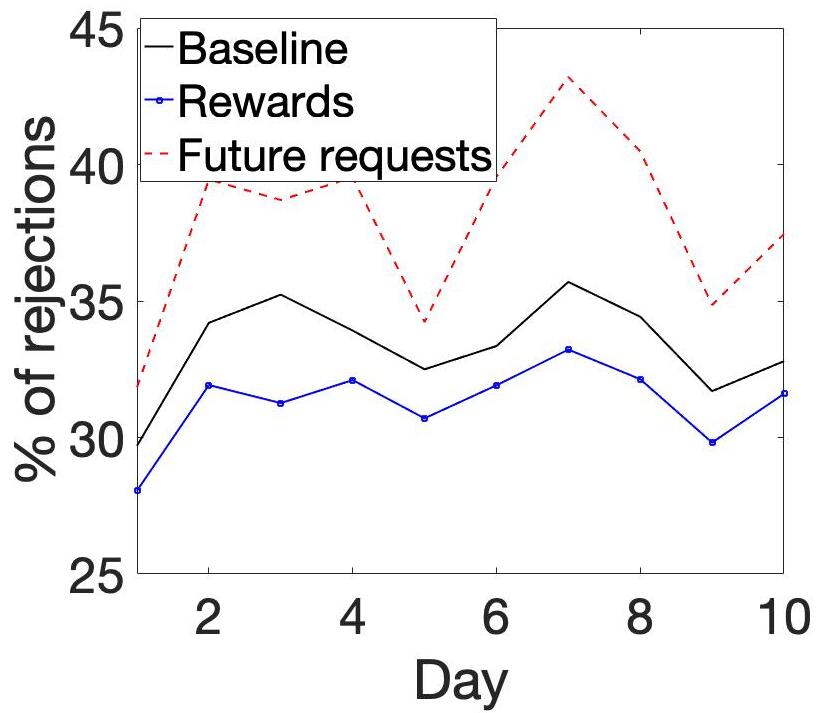} }}
   \qquad
\subfloat{{\includegraphics[width=4.8cm]{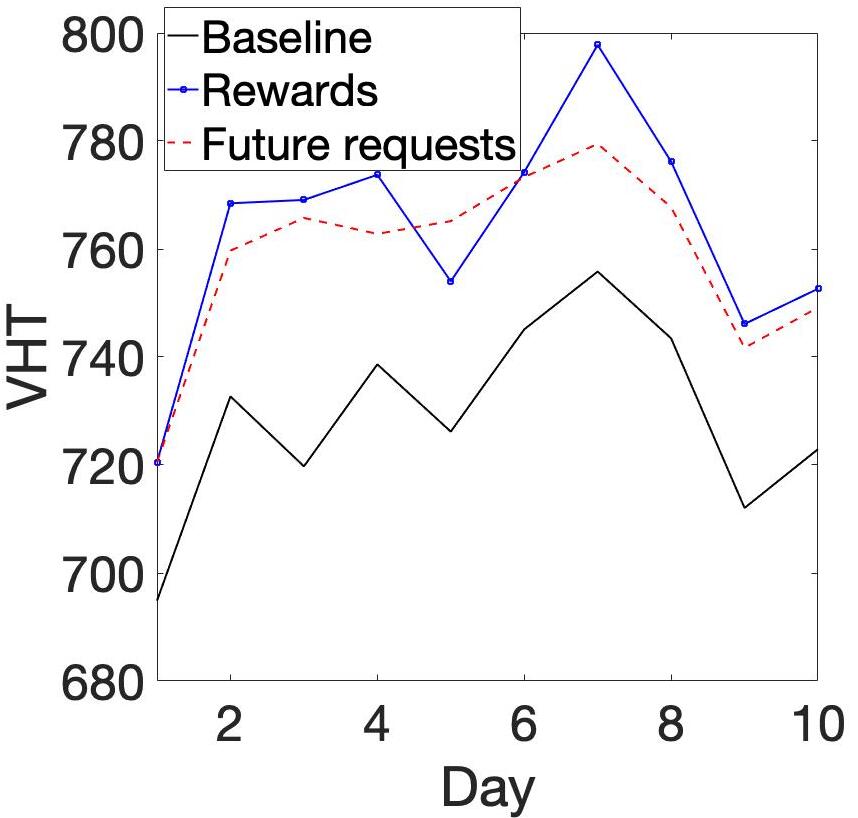} }}
   
\caption{Rejection rates (left) and vehicle-hours-traveled (right) induced by the two anticipatory methods and the baseline, when running the methods over ten different days.}
\label{Img:RejVHT10Days}
\end{figure}

\begin{figure}[H]
    \centering
\qquad
\subfloat{{\includegraphics[width=5.5cm]{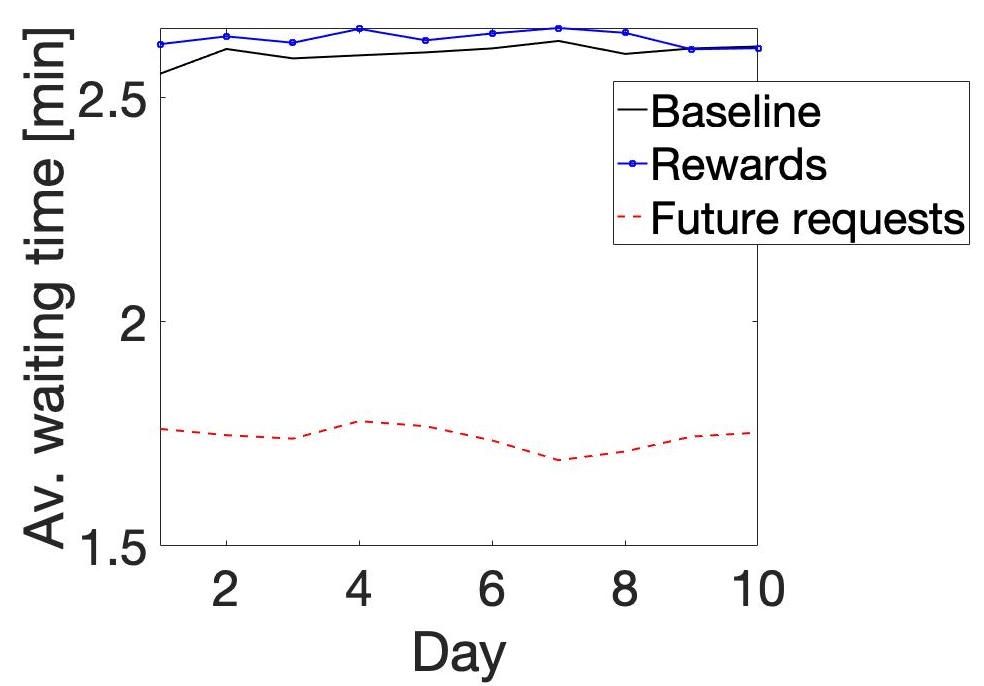}} }
   \qquad
\subfloat{{\includegraphics[width=5cm]{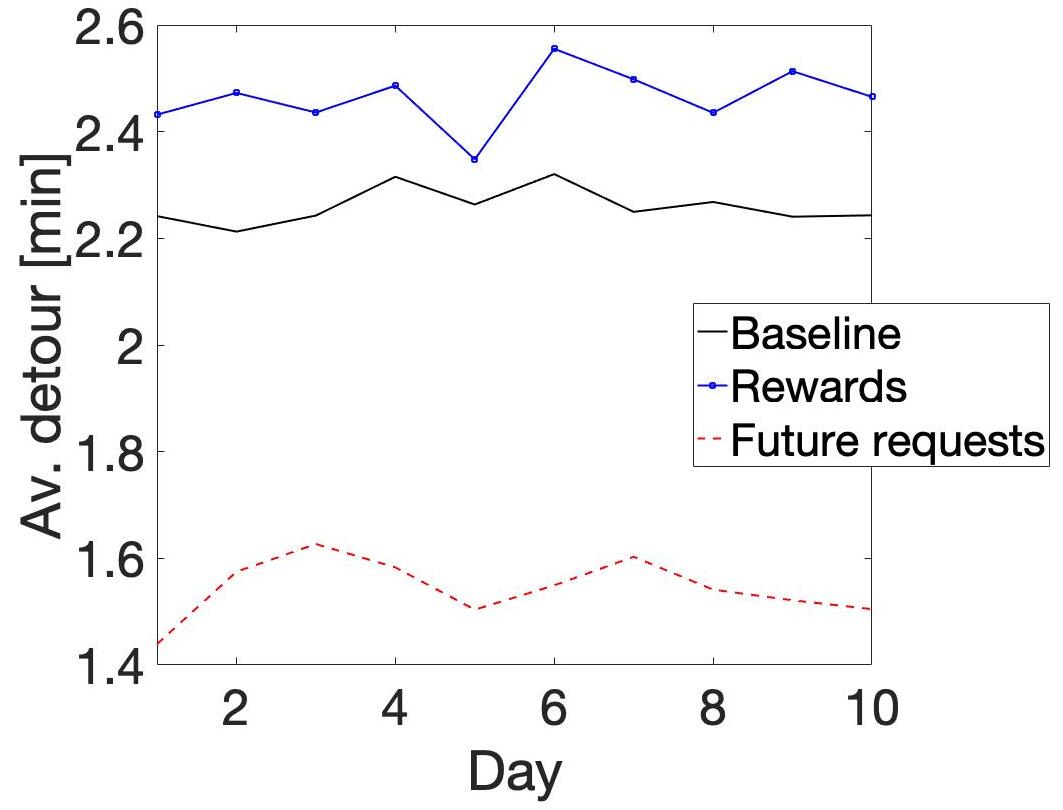}} }
   
\caption{Average users' waiting times (left) and detour (right) induced by the two anticipatory methods and the baseline, when running the methods over ten different days.}
\label{Img:TWDet10Days}
\end{figure}

Therefore, an operator should decide between using only one of the two methods. A detailed comparison among them is offered in Figures \ref{Img:RejVHT10Days}-\ref{Img:TWDet10Days}, in which we show the results when running the methods for the real requests from ten consecutive weekdays, always between 1-2 p.m. We compare the baseline (no anticipatory techniques), and the best versions of each of the two methods: $Gen_B$ with $\Theta=6$ when introducing rewards, and $Gen_B$ with $\Gamma=\frac{1}{80}$ when inserting future artificial requests. 

Figures \ref{Img:RejVHT10Days}-\ref{Img:TWDet10Days} synthesize the most relevant conclusions of the simulations. Both methods effectively impact the performance of the system, achieving different goals: whereas introducing rewards decreases the rejection rate of the system, including artificial future requests improves the quality of service for those requests that get served, and both methods require an increase in VHT. Moreover, these conclusions are robust, as they are shown to be valid for each of the ten days.

In general, introducing rewards is better for most situations, as serving more passengers is usually the most relevant purpose, even more recalling that waiting times and delay are always bounded. This method also has the virtue of not increasing the computational times at all. However, if the operator is most interested in improving the service for those passengers that are being served, inserting artificial requests is the best option.

\subsection{Detailed analysis of the impact over the system} 
\label{scn:whatishapp}
So far, we have analyzed the methods in terms of the most relevant indices of the system. However, when we introduced the need for this type of method, we justified it by analyzing the spatial heterogeneity of the results, and the influence of deciding without knowing the demand beforehand. Therefore, we now turn to analyze how the operation of the system is being modified. We focus on the method that introduces rewards, as it yields the best results, considering the rates $Gen_B$ and $Rej_{PF}$ that proved stronger.

\subsubsection{Impact over the temporal evolution of the system}

\begin{figure}[h]
    \centering
    \includegraphics[width=6cm]{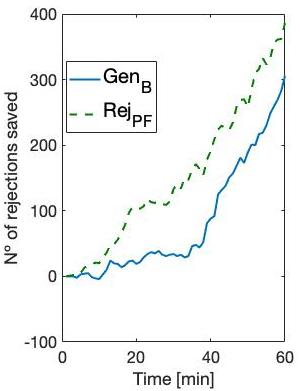}
    \caption{Difference in the accumulated rejections with or without anticipatory methods. The $y$-axis shows the difference between the number of accumulated rejections if using no anticipatory methods at all, or introducing rewards and using $Gen_B$ (solid blue curve) or $Rej_{PF}$ (dotted green curve); the $x$-axis represents the number of iterations (evolution in time), where we run one iteration per minute.}
    \label{fig:RejInTime}
\end{figure}

We first study how the number of rejections evolves in time, compared to the case with no anticipatory methods ($\Theta=0$). We know that $Gen_B$ and $Rej_{PF}$ have less rejection in total, but when does this happen? Figure \ref{fig:RejInTime} shows how many rejections are being saved thanks to introducing rewards. This is done by depicting the difference between the accumulated rejections with no rewards, and the accumulated rejections with rewards for both rates: as before, the solid blue curve represents $Gen_B$, and the dotted green curve represents $Rej_{PF}$. 

Both curves begin quite flat and even take negative values, meaning that in the first iterations, rewards worsen the system's quality. The $Rej_{PF}$'s curve rapidly starts to increase (i.e., to have fewer rejections than the method with no anticipation), whereas $Gen_B$ requires almost half an hour to do so, which verifies that the central motivation of these methods is achieved: modifying its current decisions to be better prepared for future requests. Note that, after about 40 minutes of operation, both methods reach an almost-linear increase, i.e., they keep saving rejections at a rate that keeps somewhat constant.

\subsubsection{Impact over the spatial mismatch between vehicles and requests}
\begin{figure}[H]
    \centering
\qquad
\subfloat{{\includegraphics[width=4.8cm]{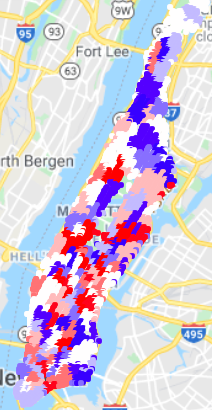}} }
   \qquad
\subfloat{{\includegraphics[width=4.8cm]{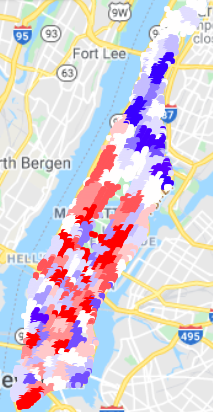}} }
\qquad
\subfloat{\includegraphics[width=1.5cm]{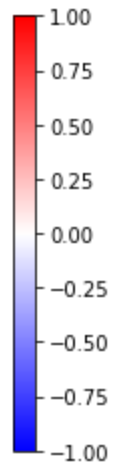}}

\caption{Differences in the location of the vehicles when using no anticipatory methods at all, or introducing rewards and using $Gen_B$ (left) or $Rej_{PF}$ (center). Each vehicle is assigned to the zone corresponding to its closest node after sixty minutes of operation of the system. A red zone means that more vehicles are assigned there in the anticipatory scheme, whereas a blue zone means the opposite; the more intense the color, the higher the difference, as shown in the colormap (right). Figures are normalized with respect to the maximum values.}
\label{Img:WhereVeh}
\end{figure}

We now analyze how the operation changes in space. We have noticed before that the most demanded zones were receiving a worse quality of service, so that more vehicles seemed to be required there. To analyze the changes, Figure \ref{Img:WhereVeh} shows, at the end of the hour that was modeled, the differences in the vehicles' positions between having no anticipatory methods and $Gen_B$ (left) or $Rej_{PF}$ (right). We partition the whole map into the same zones used for the methods with particle filters and historical data, and each vehicle is assigned to the zone corresponding to its closest node. A red sector means that there were more vehicles assigned with rewards, whereas blue means the opposite: the more intense the color, the higher the difference. 

Both Figures have almost only blue zones at the north of the network. In the center, intense red zones clearly dominate for $Rej_{PF}$, and not so clearly for $Gen_B$. That is to say, rewards are indeed moving some vehicles from the north of the network (a low demand area) to the center. It is worth saying that $Gen_B$ makes no noticeable difference in the southwest (at the bottom of the network), while $Rej_{PF}$ increases the number of vehicles there.

\begin{figure}[H]
    \centering
\qquad
\subfloat{{\includegraphics[width=4.6cm]{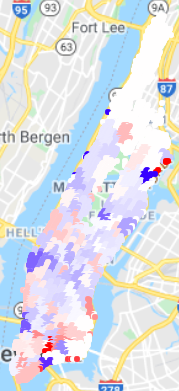}} }
   \qquad
\subfloat{{\includegraphics[width=4.7cm]{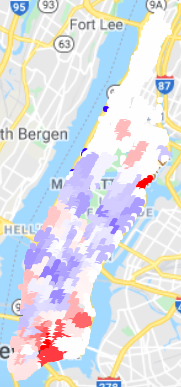}} }
\qquad
\subfloat{\includegraphics[width=1.5cm]{A simple predictive routing 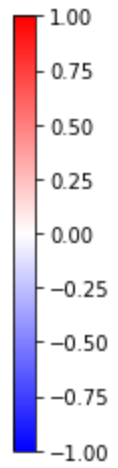}}
\caption{Differences in the spatial distribution of the rejection rates when using no anticipatory methods at all, or introducing rewards and using $Gen_B$ (left) or $Rej_{PF}$ (center). For each zone, we calculate the change in the percentage of the requests departing from there that were rejected by the system. A red zone means that more requests are rejected in the anticipatory scheme, whereas a blue zone means the opposite; the more intense the color, the higher the difference, as shown in the colorbar (right).}
\label{Img:WhereRej}
\end{figure}

A similar analysis can be done to see where the rejects are concentrated. In Figure \ref{Img:WhereRej}, each zone is an origin, that is blue if the percentage of rejected requests emerging there was higher with no rewards than with $Gen_B$ (left) or $Rej_{PF}$ (right), and red in the opposite case. Again, the more intense color, the higher the difference. 
The reduction of the rejections rates in the central zones is apparent. Notably, we see that the overall reduction is achieved by increasing the number of rejections in some zones, mostly at the south and north of the network.

When describing methods that insert rewards, a possible drawback was identified: as there is no global control of the effects introduced by the rewards, many vehicles could be sent towards the same places, which could yield a worse mismatch between vehicles and requests than with no anticipatory methods. To analyze if this is the case, we create the following indices per zone $z$, also evaluated after 60 minutes. For each zone $z$, define $v_z$ as the proportion of the vehicles inside $z$, and $r_z$ as the proportion of current requests departing from $z$. A perfectly balanced system would require that $v_z \approx r_z \forall z$. Therefore, a measure for the level of mismatch at $z$ is:

\begin{equation}
    M_z=|v_z-r_z|
\end{equation}

We choose these absolute values rather than normalized so that high-demand zones weigh more when calculating average values of $M_z$.

We now study how these values $M_z$ change when anticipatory methods are used. We use $Rej_{PF}$ for the comparison because it presents the largest impact on the system. A first analysis can be done by looking at the mean and median values of $M_z$, which is shown in Table \ref{tab:Mismatch}. As apparent, the level of mismatch is reduced when using the anticipatory scheme. Note that the absolute numbers of $M_z$ depend on many different structural aspects of the problem (such as fleet size, number of shareable requests, distances), so the relevant insight we obtain from Table \ref{tab:Mismatch} is how the numbers change when the anticipatory method is included. It is also noteworthy that $M_z$ might not take other relevant features into account (for instance, how shareable are the requests emerging from each zone), so the analysis is meaningful at a global scale rather than for specific conclusions.

\begin{table}[h]
    \centering
    \begin{tabular}{|c|c|c|}
    \hline
    \textbf{Methods} & \textbf{Mean($M_z$)} & \textbf{Median($M_z$)}  \\
    \hline 
    No rewards     & 0.0051 & 0.0035 \\
    Reward given by $Rej_{PF}$ & 0.0043 & 0.0028\\
    \hline
    \end{tabular}
    \caption{Comparison of the mean and median of the level of mismatch per zone, whith and without anticipatory methods.}
    \label{tab:Mismatch}
\end{table}

\begin{figure}[H]
    \centering
\qquad
\subfloat{{\includegraphics[width=4.6cm]{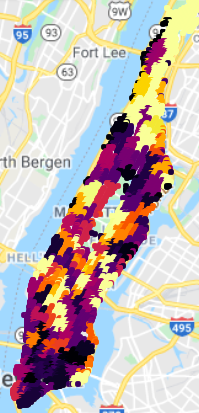}} }
   \qquad
\subfloat{{\includegraphics[width=4.6cm]{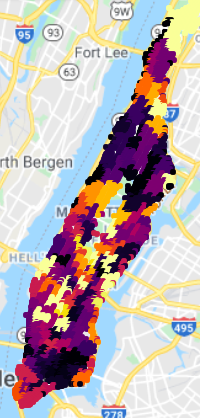}} }
\qquad
\subfloat{\includegraphics[width=1cm]{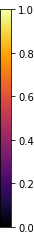}}
\caption{Spatial distribution of the level of mismatch without anticipatory methods (left), and introducing rewards given by $Rej_{PF}$ (center). Brighter colors reveal a higher level of mismatch, as shown in the colorbar (right). Values are normalized}
\label{Img:Mismatch}
\end{figure}

Conclusions from Table \ref{tab:Mismatch} are reinforced by looking at Figure \ref{Img:Mismatch}, in which we show the level of mismatch at each zone, without anticipatory methods (left), and using the rewards technique with $Rej_{PF}$ (center). Introducing the rewards builds a figure with more dark zones, i.e., more zones with a low level of mismatch. The difference is mostly achieved at the north of the network. Recalling that the rewards remove vehicles from that area, we conclude that an oversupply is now prevented there.

In all, this subsection allows us to synthesize the impact of the rewards over the ridepooling system: \textbf{using anticipatory rewards improves the system only after some initial lapse of time; the improvement is achieved by moving vehicles towards the high-demand zones, which decreases the quality of service in the low-demand zones but achieves better overall results}. Moving these cars decreases the mismatch between vehicles and requests.



\subsection{Sensibility analysis} \label{Sctn:sensibility}

So far, we have shown results for a particular network, for a fixed demand and a given fleet. We now show that the proposed techniques are robust, i.e., that they also work properly under different conditions, although results are not exactly equal. In particular, the optimal values for $\Theta$ and $\Gamma$ are highly sensitive to the specific conditions.  

Two alternative scenarios will be used for the sensibility analysis. In the first one, we serve the same demand and the same network, but using a fleet of 2000 vehicles of capacity 4, to study whether the methods are effective when the number of rejections is lower. 

\begin{figure}[H]
    \centering
    \includegraphics[width=4cm]{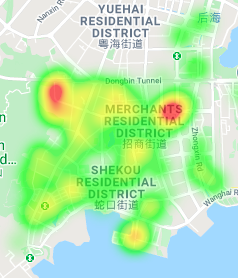}
    \caption{Heatmap of the origins of the requests in Shenzhen, China.}
    \label{fig:Shenzhen}
\end{figure}

The second scenario uses data from Didi Udian, a real ridepooling company in Shenzhen, China. As on-demand ridepooling is an emerging mobility system, the scale is much smaller: 311 requests during 12 hours. The simulation considers a fleet of 10 vehicles of capacity 5. The network is formed by 5,502 arcs and 2,201 nodes, clustered into 96 zones. A heatmap exhibiting the location of the origins is shown in Figure \ref{fig:Shenzhen}, showing that this case is of smaller case and presents a different demand structure, with two separated high-demand zones.

\subsubsection{Sensibility of the method that introduces rewards}

\begin{figure}[h]

    \centering
    \qquad
\subfloat{{\includegraphics[width=4.3cm]{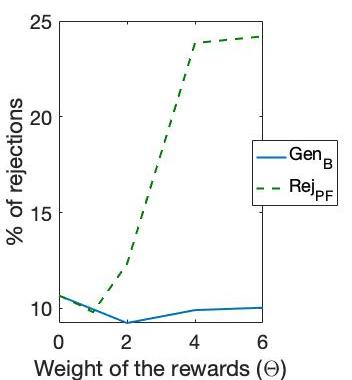}} }
   \qquad
\subfloat{{\includegraphics[width=4.4cm]{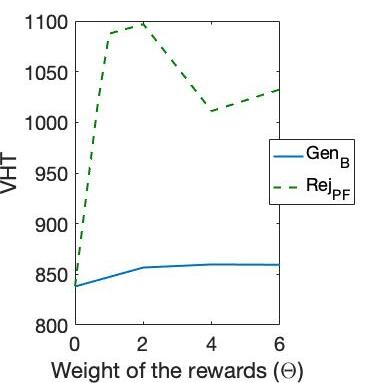}} }
\caption{Rejection rate (left) and vehicles-hour-traveled (right) for $Gen_B$ and $Rej_{PF}$, as a function of the parameter $\Theta$, when more and larger vehicles are used compared with the base scenario. $\Theta=0$ represents no anticipatory methods.}
\label{Img:SensNYC}
\end{figure}

Let us begin analyzing the method that introduces rewards. We focus on the two rates that presented the best results in the previous sections: $Gen_B$ and $Rej_{PF}$. Figure \ref{Img:SensNYC} shows the results in the same Manhattan case with a larger fleet of vehicles with a greater capacity. The most relevant conclusions are:
\begin{itemize}
    \item Introducing rewards can reduce the number of rejections even when this number is already low. The relative decrease is similar (about 10\% of the original rejections), which means that the absolute decrease is less significant.
    \item The best value for the parameter $\Theta$, however, changes a lot. In the original scenario $Gen_B$ should select $\Theta=6$ and $Rej_{PF}$ should select $\Theta=2$, and now these numbers turn to $\Theta=2$ and $\Theta=2/3$, respectively. Using a smaller $\Theta$ means that the rewards are less weighted in the objective function, which is reasonable as more vehicles can provide better results at the current time. A relevant question for future research emerges: how to determine the optimal $\Theta$ according to the external and internal conditions of the system. Note that an erroneous selection for $\Theta$ can lead to awful results, as shown by the $Rej_{PF}$ curve when $\Theta \geq 2$.
    \item $Gen_B$ presents much better results than $Rej_{PF}$ in this context, as it increases VHT just mildly.
\end{itemize}

Shenzhen's case presents several differences. The low amount of requests and vehicles allows us to analyze whether the introducing-rewards methods depend on some scale effects to be effective. 

The small scale permits describing the results in a very synthetic way and are identical for $Gen_B$ and $Rej_{PF}$: the rejection rate drops from 40.8\% to 37.3\% when $\Theta$ reaches 6 (there is no effect for lower $\Theta$), with VHT increasing from 26.9 to 27.6. That is, the results show that the methodology is sound, reducing rejections also in this scenario. 

\subsubsection{Sensibility of the method that includes future artificial requests}
\begin{figure}[h]

    \centering
    \qquad
\subfloat{{\includegraphics[width=4.5cm]{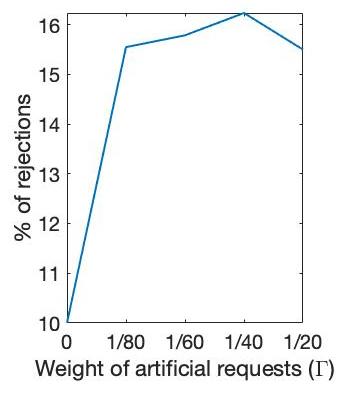}} }
   \qquad
\subfloat{{\includegraphics[width=4.5cm]{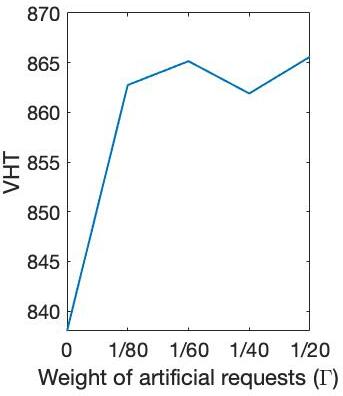}} }
\caption{Rejection rate (left) and vehicles-hour-traveled (right) when inserting future requests according to $Gen_B$, as a function of the parameter $\Gamma$, when more and larger vehicles are used compared with the base scenario. $\Gamma=0$ represents no anticipatory methods.}
\label{Img:SensFuture1}
\end{figure}

\begin{figure}[h]

    \centering
    \qquad
\subfloat{{\includegraphics[width=4.5cm]{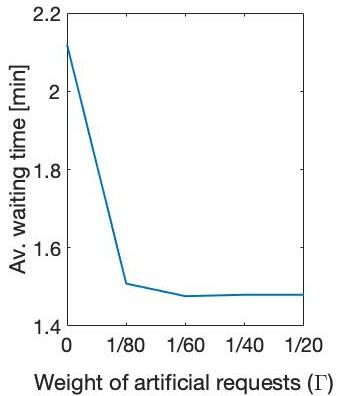}} }
   \qquad
\subfloat{{\includegraphics[width=4.5cm]{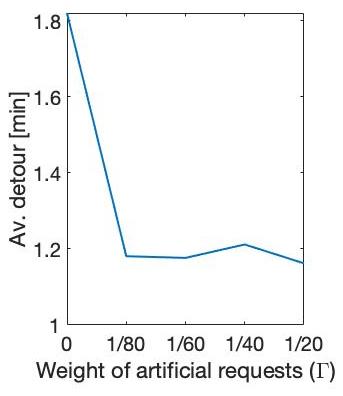}} }
\caption{Average waiting time (left) and detour (right) when inserting future requests according to $Gen_B$, as a function of the parameter $\Gamma$, when more and larger vehicles are used compared with the base scenario. $\Gamma=0$ represents no anticipatory methods.}
\label{Img:SensFuture2}
\end{figure}

To analyze the robustness of this method, we consider the same two scenarios. We use the basic generation rate because it has presented the best results so far. When applied over Manhattan with a greater fleet of larger vehicles, results are exhibited in Figures \ref{Img:SensFuture1} and \ref{Img:SensFuture2}. Results are consistent with the base scenario: this method allows for a reduction in waiting times and detours at the cost of increasing the number of rejections and VHT. However, the relative losses regarding rejection rates are worse here than in the base scenario, and the gains are similar, i.e., this method is less effective when using a larger fleet.

\begin{figure}[H]
    \centering
    \qquad
\subfloat{{\includegraphics[width=4.65cm]{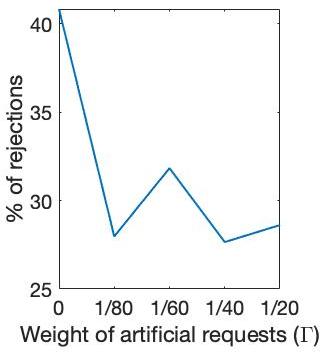}} }
   \qquad
\subfloat{{\includegraphics[width=4.5cm]{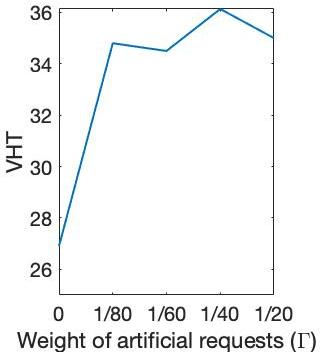}} }
\caption{Rejection rate (left) and vehicles-hour-traveled (right) when inserting future requests according to $Gen_B$, as a function of the parameter $\Gamma$, in the Shenzhen study case. $\Gamma=0$ represents no anticipatory methods.}
\label{Img:ShenzenFuture1}
\end{figure}

\begin{figure}[H]
    \centering
    \qquad
\subfloat{{\includegraphics[width=4.55cm]{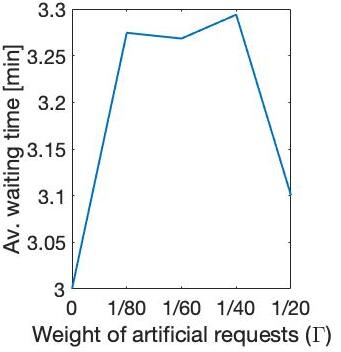} }}
   \qquad
\subfloat{{\includegraphics[width=4.5cm]{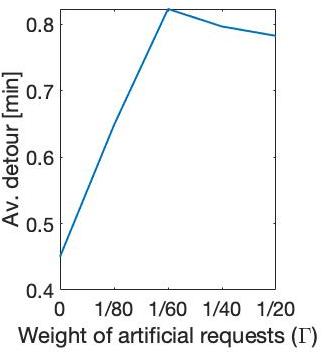} }}
\caption{Average waiting time (left) and detour (right) when inserting future requests according to $Gen_B$, as a function of the parameter $\Gamma$, in the Shenzhen study case. $\Gamma=0$ represents no anticipatory methods.}
\label{Img:ShenzenFuture2}
\end{figure}

However, when solving the Shenzhen case (Figures \ref{Img:ShenzenFuture1} and \ref{Img:ShenzenFuture2}), results are inverted compared with the base scenario: the rejection rate drops from 40.8\% to 27.7\% in the most extreme case ($\Gamma=\frac{1}{60}$), while average waiting time increases from 3 [min] to 3.29 [min] average detours from 0.45 [min] to 0.8 [min], and VHT increases from 26.9 to 36.1. 

This situation might look paradoxical but can be explained. Let us first say that in this case, we only insert one future request per iteration ($m=1$) to preserve a proportion between real and artificial requests that is somewhat similar to the one used in Manhattan (where total future requests represent between a third and a half of total real requests). Nevertheless, the low number of real requests in Shenzhen implies that several iterations have none, meaning that the artificial requests are not competing with the real ones on those occasions; recall that such a competition was identified as the explanation of why the rejection rate could increase with this anticipatory technique. Moreover, no heuristic is needed in this case, which was the other crucial factor in explaining the increase in the rejection rate for the basic scenario.

In conclusion, we can state that this method outperforms the one that introduces rewards when the scale of the problem is small. Moreover, in the small scale case, the computational time (that was said to be a relevant drawback of this method) is not an issue at all.

\section{Conclusions and future research} \label{scn:conclusionsAnt7}
In this paper, we propose techniques to reduce the mismatch between vehicles' positions and users' origins that tends to occur in on-demand ridepooling systems (when the assignment decisions do not take the future into account). Our techniques only require the information that is generated when operating the system, by looking at the zones where most requests are emerging and being rejected.

We propose two anticipatory techniques. In the first one, we modify the objective cost function by introducing a reward (a negative additive term) to each feasible matching between a vehicle and a group of requests, such that the reward increases when the vehicle's route finishes in a high-demand area. This technique acts at an individual level, i.e., it does not include control of the system's operation as a whole, which permits not to increase the computational burden. The second technique includes artificial future requests emerging from the high-demand zones, to be assigned together with the real current ones, so that if a vehicle is assigned to a future request, it will be moved towards its origin. This technique yields a system-wide decision, at the cost of increasing the computational time, that might require including other heuristics to compensate.

How ``high-demand'' an area is, is defined as a function of the number of requests emerging there (generation rates), or of the number of rejected requests that departed from there (rejection rates). We propose three ways of calculating the 
rates of a node: i) the direct number of users departing from it, ii) a spatial smoothing of that direct number,  and iii) a temporal stabilization of that direct number through a particle filter method. The last one requires aggregating the nodes into zones. 

These approaches are applied over the assignment method proposed by \cite{alonso-mora_-demand_2017}, using Manhattan as a study case. Results reveal that introducing rewards effectively reduces the number of rejections, at the cost of increasing total delay for the served users. In contrast, the inclusion of artificial future requests reduces total delay but increasing the number of rejections unless the number of requests is small enough to have zero real requests during some minutes. Both methods require vehicles to move more. Results depend on the rates being used, and applying both methods together does not yield good results.

A detailed analysis of the impact of the rewards method over the system's operation reveals that the immediate effect can be negative, but after some minutes, the system begins to show an enhanced performance, which is expected as these are anticipatory techniques. Vehicles are shown to be moved from low to high-demand areas, which might degrade the qualithy of service for users departing from low-demand nodes, but the improvement in the high-demand zones outperforms this worse quality of service.

As on-demand ridepooling is an emerging mobility system, there is plenty of room for future research. A relevant direction identified through the paper is the optimal selection of the parameters $\Theta$ and $\Gamma$, i.e., the relative weight of the anticipatory components. Different methods and scenarios require adapting $\Theta$ and $\Gamma$, so using a non-constant value would be ideal. How to tune this value according to the external and internal conditions is a challenging and relevant question that could be addressed using learning procedures. Additional research questions deal with the relationship between these techniques and rebalancing methods: is it possible to adapt the rebalancers to take better advantage of the anticipatory techniques, and vice-versa? 

\section*{Acknowledgements}
We thank the valuable comments by Dr. Xiaoshan Bai, from Shenzhen University. This research was partially funded by Didi Udian Technology (Shenzhen) Co. Ltd.

\section*{Authors' contribution}
All authors contributed to the study conception and design. Numerical experiments were run by Andres Fielbaum. The first draft of the manuscript was written by Andres Fielbaum and all authors commented on previous versions of the manuscript. All authors read and approved the final manuscript.

\section*{Conflict of interest}
On behalf of all authors, the corresponding author states that there is no conflict of interest.

\bibliographystyle{apalike}
\bibliography{bib}

\begin{tiny}
\begin{table*}[]
    \centering
    \begin{tabular}{|c|c|c|}
    \hline
    \textbf{Symbol} & \textbf{Meaning} & \textbf{Numeric value} (if applies)  \\
    \hline
    $G=(V,E)$ & Graph representing the network & -\\
    $t_V(u,w)$ & Shortest time required to go from node $u$ to node $w$ & - \\
    $PO$ & Length of the period of operation & 1 [h] \\
    $\delta $ & Time-lapse between each assignment process (receding horizon) & 1 [min]\\
    $r$ & A single request & - \\
    $tr_r$ & Time at which $r$ emerges & - \\
    $o_r,d_r$ & Origin and destination of $r$ & - \\
    $\mathcal{R}_{all}$ & Set of requests emerged throughout the operation period & - \\
    $\mathcal{R}_s(t)$ & Requests that are being served at time $t$ & - \\
    $\mathcal{R}_e(t)$ & Requests that are waiting to be assigned at time $t$ & - \\
    $\mathcal{R}_c(t)$ & Requests that were dropped off before $t$ & - \\
    $\mathcal{R}_r(t)$ & Requests that were rejected before $t$ & - \\
    $\mathcal{V}(t)$ & State of the fleet of vehicles at time $t$ & - \\
    $v$ & A single vehicle & - \\
    $Pos_v(t)$ & Position of $v$ at time $t$ & - \\
    $Req_v(t)$ & Requests being served by $v$ at time $t$ & - \\
    $\pi_v(t)$ & Planned route of $v$ at time $t$ & - \\
    $T$ & One trip (set of requests) & - \\
    $\mathcal{C}$ & Constraints over feasible assignments & - \\
    $c(v,T,\pi,t)$ & Cost of assigning $T$ to $v$, with updated route $\pi$, at time $t$ & - \\
    $c_U,c_O$ & Users' and operator's costs & - \\
    $\Psi=\{\tau_1,...,\tau_N\}$ & Instants in which an assignment is decided & - \\
    $A$ & Assignment between vehicles and requests & - \\
    $Q(\tau_i,A)$ & Set of requests rejected by $A$ & - \\
    $c_R(Q(\tau_i,A))$ & Cost of rejecting $Q(\tau_i,A)$ & - \\
    $\mathcal{A}(\mathcal{R}_{e}(\tau_i),\mathcal{V}(\tau_i),\mathcal{C})$ & Set of feasible assignments between $\mathcal{R}_{e}(\tau_i)$ and $\mathcal{V}(\tau_i)$ fulfilling $\mathcal{C}$ & - \\
    $\mathcal{R}_{ok}, \mathcal{R}_{ko}$ & Requests served and rejected throughout the operation period & - \\
    $\Pi_v$ & Route followed by $v$ throughout the operation period & - \\
    $c_A$ & Cost function with rewards & - \\
    $p_w$ & User's cost of waiting one unit of time & 4.64 [\$US/h] \\
    $t_w$ & Waiting time & - \\
    $p_v$ & User's cost of spending one unit of time over the vehicle & 2.32 [\$US/h] \\ 
    $D$ & Detour & - \\
    $p_O$ & Operator's cost of moving a vehicle one unit of time & 3.48 [\$US/h] \\
    $p_{KO}$ & Cost of rejecting a request & 3.09 [\$US] \\
    $L$ & Length of a route & - \\
    $R$ & Reward function & - \\
    $\Theta$ & Tuning parameter for reward functions & 0-6 \\
    $LN(\pi)$ & Last node of route $\pi$ & - \\
    $IN(v,T,\pi)$ & First node in $\pi$ such that the vehicle is not full from there on  & -\\
    $m$ & Number of artificial future requests & 50 \\
    $\Gamma$ & Ratio between rejection rates of real and artificial requests & $\in (0,1)$\\
    $\phi$ & Time between artificial request times & 1 [min] \\
    $Gen_B(u,\tau_i)$ & Basic generation rate of node $u$ at time $\tau_i$ & -\\
    $Rej_B(u,\tau_i)$ & Basic rejection rate of node $u$ at time $\tau_i$ & -\\
    $Gen_S(u,\tau_i)$ & Smooth generation rate of node $u$ at time $\tau_i$ & -\\
    $Rej_S(u,\tau_i)$ & Smooth rejection rate of node $u$ at time $\tau_i$ & -\\
    $Gen_{PF}(u,\tau_i)$ & Generation rate of node $u$ at time $\tau_i$, based on particle filters & -\\
    $Rej_{PF}(u,\tau_i)$ & Rejection rate of node $u$ at time $\tau_i$, based on particle filters & -\\
    $Gen_H(u,\tau_i)$ & Generation rate of node $u$ at time $\tau_i$, based on historical data & -\\
    $\psi$ & Tuning parameter of the smooth method & 1 \\
    $M$ & Number of zones in the network & 167 \\
    $t_M$ & Upper bound to the distance between a node and the center of its zone & 150 [s]\\ 
    $\eta$ & Number of samples for the Montecarlo method & 100 \\
    $\lambda_{zi}$ & Auxiliary variables for the particle filter method & - \\
    $w_{zi}$ & Weights for the particle filter method & - \\
    $\sigma^2$ & Volatility parameter in the particle filter method & 0.05 \\
    $\mathcal{D}$ & Number of the days considered for the historical dataset & 9 \\

    \hline
    \end{tabular}
    \caption{Glossary of symbols used throughout the paper, and the numeric value of the parameters in the models.}
    \label{tab:my_label}
\end{table*}

\end{tiny}
\end{document}